\documentclass[12pt,a4paper]{article}
\usepackage{macros}
\usepackage{amssymb}

 \def\a{\alpha}

\preprint{}
\title{Connection formulae in the Collision Limit I: Case Studies in Lifshitz Geometry}

\author[a]{Hao Zhao}
\author[a,b]{Rui-Dong Zhu}
\affiliation[a]{School of Physical Science and Technology, Soochow University, 333 Ganjiang Road, 215006 Suzhou, China
}
\affiliation[b]{Institute for Advanced Study, Soochow University, 333 Ganjiang Road, 215006 Suzhou, China}
\emailAdd{hzhao001@stu.suda.edu.cn,rdzhu@suda.edu.cn}

\abstract{The connection formulae provide a systematic way to compute physical quantities, such as the quasinormal modes, Green functions, in blackhole perturbation theories. In this work, we test whether it is possible to consistently take the collision limit, which brings two or more regular singularities into an irregular one, of the connection formulae, and we provide some supportive evidence for it. }

\begin{document}

\allowdisplaybreaks

\maketitle

\section{Introduction}

Exact results of non-perturbative effects in supersymmetric gauge theories have been thought to be rather irrelevant to blackhole physics before the interesting proposal of \cite{Aminov:2020yma} that Nekrasov instanton partition functions of 4d $\cN=2$ gauge theories can be used to compute physical quantities such as the quasinormal modes (QNMs) in the perturbation theory around blackholes. There have been intensive studies on the relevant topics \cite{Hatsuda:2020sbn,Hatsuda:2020iql,Bonelli:2021uvf,daCunha:2021jkm,Bianchi:2021mft,Nakajima:2021yfz,Hatsuda:2021gtn,Fioravanti:2021dce,Bonelli:2022ten,Bianchi:2022wku,Dodelson:2022yvn,Consoli:2022eey,Imaizumi:2022qbi,Fioravanti:2022bqf,Lisovyy:2022flm,Bhatta:2022wga,daCunha:2022ewy,Gregori:2022xks,Imaizumi:2022dgj,Bianchi:2022qph,Fioravanti:2023zgi,Cano:2023tmv,Bianchi:2023rlt,Fucito:2023plp,Giusto:2023awo,Aminov:2023jve,Kwon:2023ghu,Hatsuda:2023geo,Bhatta:2023qcl,He:2023wcs,Lei:2023mqx,BarraganAmado:2023wxt,Fucito:2023afe,Bautista:2023sdf,DiRusso:2024hmd,Ge:2024jdx} since then, and similar techniques are now applicable to QNMs, tidal deformations, the greybody factor, and the inspiral waveform etc. related to the gravitational wave physics of various types of blackhole models. The reason behind this connection is simple. The perturbation theory in a blackhole background is characterized by a (usually second-order) linear differential equation with polynomial coefficients, 
\begin{equation}
    \lt(A(z)\frac{{\rm d}^2}{{\rm d}z^2}+B(z)\frac{{\rm d}}{{\rm d}z}+C(z)\rt)\psi(z)=0,\label{diff-eq}
\end{equation}
where $A(z)$, $B(z)$ and $C(z)$ are polynomials of $z$. Such linear differential equations can be classified by the singularities in the equation. This can be clearly seen after we bring the differential equation to its Schr\"odinger form, 
\begin{equation}
    \lt[\frac{{\rm d}^2}{{\rm d}z^2}+Q(z)\rt]\Psi(z)=0,
\end{equation}
where 
\begin{equation}
    Q(z)=\frac{C(z)}{A(z)}-\frac{B^2(z)}{4A^2(z)}+\lt(-\frac{B(z)}{2A(z)}\rt)',
\end{equation}
and 
\begin{equation}
    \Psi(z)=\psi(z)\exp\lt(\int^z{\rm d}x\frac{B(x)}{2A(x)}\rt).
\end{equation}
The singularities appear in the potential $Q(z)$ as divergent points $z_i$ of the form 
\begin{equation}
    \frac{b_i}{(z-z_i)^k}\subset Q(z), 
\end{equation}
and they are usually classified into regular singularities (with $k\leq 2$) and irregular ones (with $k>2$). Correspondingly 4d $\cN=2$ supersymmetric gauge theories have class ${\cal S}$ construction in the context of string theory \cite{Gaiotto:2009we}. In this construction, one compactifies 6d $\cN=(2,0)$ theory on a punctured Riemann surface $\Sigma$ to obtain 4d supersymmetric gauge theories. The Riemann surface $\Sigma$ gives the well-known Seiberg-Wittern curve \cite{Seiberg:1994rs,Seiberg:1994aj}, which describes the low-energy effective theory of 4d $\cN=2$ gauge theory. By further considering the Nekrasov-Shatashvili limit of such gauge theories \cite{Nekrasov:2009rc}, the Seiberg-Witten curve is then quantized to a linear differential equation. Then a one-to-one correspondence can be established between the singularities in the linear differential equation and the punctures on $\Sigma$. That is, given a second-order linear differential equation \eqref{diff-eq}, one can always find a corresponding $\cN=2$ theory. 

To solve for the QNMs and other related physical quantities, we usually need to solve the following boundary condition problem: consider a second-order linear differential equation on an interval $[z_0,z_1]$ between two singularities $z_0$ and $z_1$ in the equation, and then impose boundary conditions on two ends of the interval to find the allowed parameter spectra of the differential equation. One of the most standard and systematic ways to solve such a boundary condition problem is to use the connection formulae. Since one can always find two independent solutions to the linear differential equation as series expansions around each singularity, $\psi_\pm^{(z_0)}(z)$ and $\psi_\pm^{(z_1)}(z)$, then if we know the relation between these two sets of solutions, for example, one set of solutions can be analytically continued to the other singularity, as
\begin{equation}
    \psi_\theta^{(z_0)}(z)=\sum_{\theta'=\pm}A_{\theta,\theta'}\psi_{\theta'}^{(z_1)}(z),\quad \theta=\pm,
\end{equation}
for some linear combination coefficients $A_{\theta,\theta'}$, the boundary condition problem will be solved straightforwardly. For example, if the solutions $\psi^{(z_0)}_+$ and $\psi^{(z_1)}_-$ respectively satisfy the boundary condition at $z=z_0$ and $z=z_1$, then we simply require 
\begin{equation}
    A_{++}=0,
\end{equation}
to solve the problem. 

In general, however, it is very difficult to derive the connection formulae, except for several simple examples such as the hypergeometric and Bessel functions, which have integral representations that allow one to write down such formulae. When all the singularities in the differential equation are regular (i.e. $k\leq 2$), the linear differential equation can be obtained from the so-called Belavin–Polyakov–Zamolodchikov (BPZ) equation satisfied by the correlation function in 2d CFT, 
\begin{equation}
    \lt(b^{-1}\frac{\partial^2}{\partial z^2}+\sum_{i=1}^n\frac{\partial_{z_i}}{z-z_i}+\frac{\Delta_i}{(z-z_i)^2}\rt)\langle V_{2,1}(z)\prod_{i=1}^nV_{i}(z_i)\rangle=0,
\end{equation}
in the semiclassical limit with the central charge $c\to \infty$. We note that this fact is consistent with the Alday-Gaiotto-Tachikawa relation \cite{Alday:2009aq,Wyllard:2009hg} between 4d $\cN=2$ theories and 2d CFTs, which states that the correlation functions in 2d CFT equal to the Nekrasov partition function of the dual gauge theory. In the case with only regular singularities, the connection formulae can be guessed from the crossing symmetry of the correlation function \cite{Bonelli:2022ten} (see also \cite{Jeong:2018qpc,Lei:2023mqx,Liu-Zhu}). When irregular singularities appear in the linear differential equation, the situation becomes more complicated and it is even more difficult to obtain the connection formulae. In the case of one irregular singularity (with $k=3$ or $k=4$) plus one or more regular singularities, the connection formulae can be obtained with the help of the known connection formulae of the Bessel and Whittaker function and the crossing symmetry of the corresponding irregular conformal blocks \cite{Bonelli:2022ten}. In the existence of the irregular singularity of $k>4$, no result has been reported in the literature. In the context of the gauge theories, for gauge theories with Lagrangian description (and thus with clear known expressions of instanton partition function), the corresponding connection formulae are rather clear, while it suddenly becomes much harder to obtain the connection formulae once we turn to the so-called non-Lagrangian theories, such as the Argyres-Douglas (AD) theories. The AD theories are theories with a strongly coupled nature \cite{Argyres:1995jj,Xie:2012hs}, and therefore there is still no known closed-form expression for the Nekrasov partition function, which causes difficulties in analysis. 

\begin{figure}
    \centering
    \includegraphics[width=10cm]{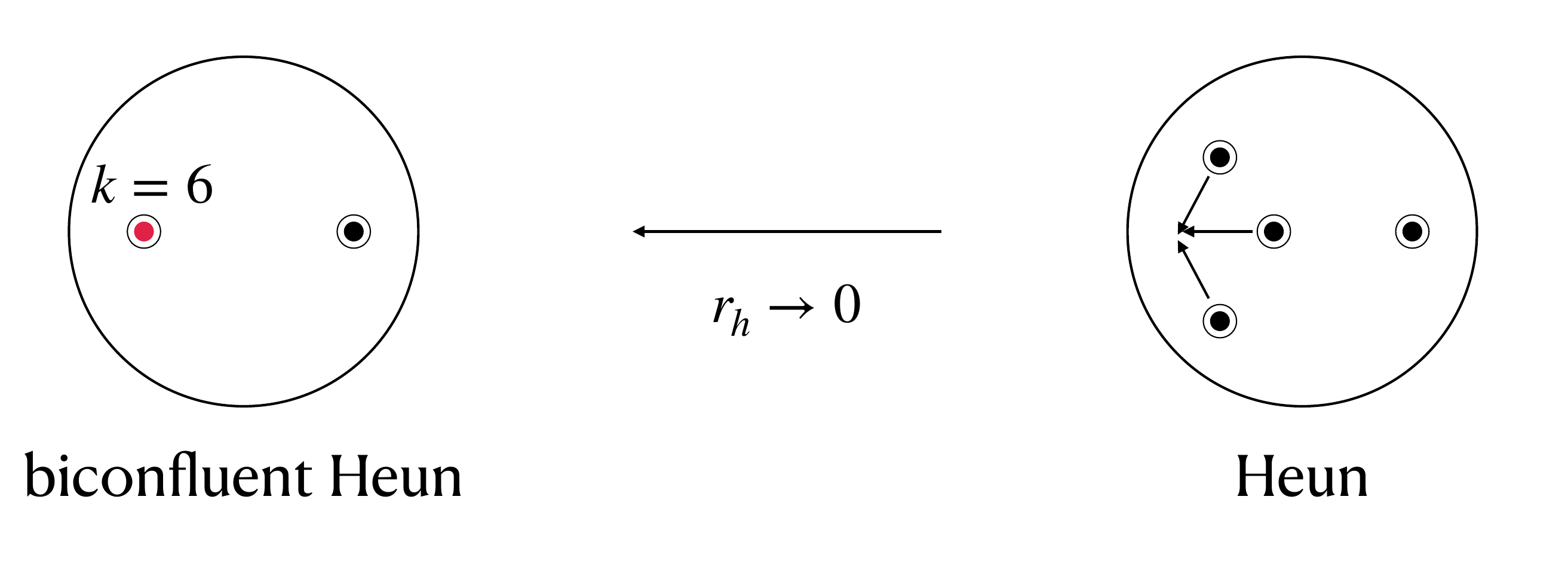}
    \caption{The Heun function reduces to the biconfluent Heun function, when three regular singularities (black dots) are merged into an irregular singularity (a red dot) of $k=6$.}
    \label{fig:HeunB}
\end{figure}

In this article, we try to explore a naive idea: to obtain the connection formulae with irregular singularities directly by taking the collision limit of the connection formulae with only regular singularities. The collision limit refers to the limit that brings two or more regular singularities to coincide, for example in the case of two regular singularities colliding into an irregular one, we have 
\begin{equation}
    \lim_{z_2\to z_1}\frac{b_1}{(z-z_1)^2}+\frac{b_2}{(z-z_2)^2}=\frac{b'_1}{(z-z_1)^k}+{\cal O}((z-z_1)^{-k+1}).
\end{equation}
The power $k$ here depends on the way we take the limit, but it cannot exceed $2n$ when $n$ regular singularities collide together. As stated above, when the differential equation has only regular singularities, the connection formulae can be obtained from 2d CFT and gauge-theory data, so it is tempting to start from a known connection formula and take its collision limit to write down a new connection formula for the differential equation with irregular singularities of $k>4$. However in the collision limit, a lot of divergent factors will appear, and in general, it is not clear whether the naive idea to directly take the limit of the connection formulae works. In this work, we test this idea and provide some supportive evidence for it by starting from cases with known connection formulae, hypergeometric functions and Heun functions, to see if the collision limit may reproduce the expected connection formulae respectively given by the Bessel and the Whittaker functions. In the first case, two regular singularities are combined into one irregular singularity of $k=3$, and in the second case, three regular singularities in the Heun equation merge into an irregular singularity of $k=6$, which characterizes the biconfluent Heun functions (see Figure \ref{fig:HeunB}). 

The article is organized as follows. In section \ref{s:L-brane} and \ref{s:L-geometry}, we review the computation of QNMs in two non-relativistic geometries, the Lisfhitz brane and the Lifshitz geometry. In particular, we show that in two cases of special parameters, the Klein-Gordon equation has known connection formulae, and taking the horizon size to zero, $r_h\to 0$, brings the Lifshitz brane to the Lifshitz geometry. This limit corresponds exactly to the collision limit mentioned before. Then in section \ref{s:wave} and \ref{s:conn}, we discuss how the wavefunctions and the connection formulae behave in the collision limit, and we aim to provide some evidence to suggest that it may be possible to take the collision limit of the connection formulae to analyze the linear differential equations with wild irregular singularities of $k>4$. 

\section{Quasinormal modes in Lifshitz branes}\label{s:L-brane}

In this section, we summarize two cases in the Lifshitz brane, in which one can use the known connection formulae to perform analytic studies on the QNMs and the Green function. In the next section, we will exhibit the corresponding computations in the Lifshitz-geometry limit. We are interested in how such a limit, where two or more singularities in the differential equation collide into one, is taken care of at the level of the connection formulae. 

The metric of the Lifshitz brane is given by 
\begin{equation}
    {\rm d}s^2=\frac{1}{r^2f^2(r)}{\rm d}r^2-f^2(r)r^{2z}{\rm d}t^2+r^2{\rm d}\vec{x}^2_{d-1},
\end{equation}
where 
\begin{equation}
    f^2(r)=1-\lt(\frac{r_h}{r}\rt)^{d+z-1},
\end{equation}
with $r_h$ denoting the location of the horizon and $r=\infty$ corresponding to the boundary (i.e. spatial infinity) of the spacetime. 
The tortoise coordinate is given by 
\begin{equation}
    {\rm d}r_\ast=\frac{1}{r^{z+1}f^2(r)}{\rm d}r,
\end{equation}
and in the Lifshitz brane, it simply integrates to a hypergeometric function, 
\begin{equation}
    r_\ast=-\frac{r^{-z}}{z}\ _2F_1\lt[1,\frac{z}{d+z-1},1+\frac{z}{d+z-1},\lt(\frac{r_h}{r}\rt)^{d+z-1}\rt].
\end{equation}

Depending on the value of the Lifshitz scaling exponent $z$ and the space dimension $d$, the QNMs can exhibit rather different features as shown in \cite{Sybesma:2015oha}. In particular, for some special combinations of $d$ and $z$, the Klein-Gordon equation obeyed by the scalar perturbations $\phi$ around the Lifshitz brane geometry, 
\begin{equation}
    \lt(\frac{\partial^2}{\partial r^2_\ast}+\omega^2-V(r)\rt)\tilde{\phi}=0,\label{KG-brane}
\end{equation}
simplifies to some differential equation of known special functions, and one may obtain analytic results on the QNMs. To obtain the Klein-Gordon equation \eqref{KG-brane}, we adopt the following separation of variables of $\phi$, 
\begin{equation}
    \phi=R(r)e^{-i\omega t+i\vec{k}\cdot\vec{x}}=r^{\frac{d-1}{2}}\tilde{\phi}(r)e^{-i\omega t+i\vec{k}\cdot\vec{x}},
\end{equation}
and for later purposes, we consider the massive Klein-Gordon equation $\Box \phi=m^2\phi$ to have  
\begin{equation}
    V(r)=r^{2z}f^2(r)\lt(\frac{k^2}{r^2}+m^2+\frac{(d-1)(d+2z-1)}{4}+\frac{(d-1)^2}{4}\lt(\frac{r_h}{r}\rt)^{d+z-1}\rt).\label{potential}
\end{equation}
The QNMs are obtained by requiring the boundary conditions as follows: at the boundary on the spacetime, i.e $r=\infty$, we impose a Dirichlet boundary condition, $\phi\rightarrow 0$; at the horizon, $r=r_h$, we require that only the ingoing modes are allowed, i.e. 
\begin{equation}
    \phi\sim e^{-i\omega (t+r_\ast)},\quad r\sim r_h,
\end{equation}
where 
\begin{equation}
    \exp\lt(-i\omega r_\ast\rt)\sim (r-r_h)^{-\frac{i\omega}{r_h^z(d+z-1)}},\quad r\sim r_h.
\end{equation}

\paragraph{Remark:} We note that the asymptotic behavior of the solution $\phi$ at the spatial infinity is completely different in the Lifshitz brane (including AdS blackholes with $z=1$) and in the asymptotic flat spacetime. In the latter case, the potential $V(r)$ vanishes at the spatial infinity and thus two solutions behave as ingoing and outgoing waves, $\phi\sim e^{\pm i\omega r_\ast}$. However in the Lifshitz brane case, the potential \eqref{potential} diverges at $r=\infty$, and two independent solutions behaves asymptotically as $\tilde{\phi}\sim r^{\frac{\pm\sqrt{z^2+4}-z}{2}}$. The physical solution in this case thus satisfies the Dirichlet boundary condition instead at the boundary. 
\begin{flushright}
    $\Box$
\end{flushright} 

In this article, we focus on two cases in which the solutions can be written in terms of known special functions, and then we investigate the collision limit $r_h\rightarrow 0$ that brings the Lifshitz brane to a so-called Lifshitz geometry. 

The first special case we consider is $d=z+1$ and $k=0$. It has been known that the solutions in this case are given by the hypergeometric functions $_2F_1$ \cite{Sybesma:2015oha}, and changing to the variable $x=\frac{r}{r_h}$, one obtains two independent solutions of $\phi$ as\footnote{We note that one should be careful that the KG equation \eqref{KG-brane} is for $\tilde{\phi}$, but the solutions presented here are for $R=r^{\frac{z}{2}}\tilde{\phi}$.} 
\begin{align}
    R_1(x)=x^{m_z-\frac{z}{2}}(-1+x^{-2z})^{\frac{i \Tilde{\omega}}{2z}}\ _2F_1\lt[\frac{-m_z+z+i\Tilde{\omega}}{2z},\frac{-m_z+z+i\Tilde{\omega}}{2z},1-\frac{m_z}{z},x^{-2z}\rt],\cr
    R_2(x)=x^{-m_z-\frac{z}{2}}(-1+x^{-2z})^{\frac{i \Tilde{\omega}}{2z}}\ _2F_1\lt[\frac{m_z+z+i\Tilde{\omega}}{2z},\frac{m_z+z+i\Tilde{\omega}}{2z},1+\frac{m_z}{z},x^{-2z}\rt].
\end{align}
where $\tilde{\omega}:=\frac{\omega}{r_h^z}$ and $m_z:=\sqrt{m^2+z^2}$. The above two solutions are around $x\sim \infty$, but we also need the solutions around $x\sim 1$. Let us further denote 
\begin{equation}
    R(x)=x^{\frac{z}{2}}(x^{2z}-1)^{\frac{i \Tilde{\omega}}{2z}}w(x^{-2z}),\label{R-to-w}
\end{equation}
then of course $w(x)$ satisfies the hypergeometric equation with the identification, 
\begin{align}
    a=\frac{-m_z+z+i\tilde{\omega}}{2z},\quad b=\frac{m_z+z+i\tilde{\omega}}{2z}=a+\frac{m_z}{z},\quad c=1.
\end{align}
Two solutions around $x=\infty$ in the hypergeometric equation are given by
\begin{align}
    &w_3(x)=(-x)^{-a}\ _2F_1[a,1+a-c,1+a-b,x^{-1}],\\
    &w_4(x)=(-x)^{-b}\ _2F_1[b,1+b-c,1+b-a,x^{-1}],
\end{align}
and around $x=1$, two independent solutions read 
\begin{align}
    &w_2(x)=\ _2F_1[a,b,1+a+b-c,1-x],\\
    &w_6(x)=(1-x)^{c-a-b}\ _2F_1[c-a,c-b,1+c-a-b,1-x].
\end{align}
One can check that $w_4$ and $w_6$ respectively satisfy the boundary conditions at the boundary $x=\infty$ and the horizon $x=1$. To impose these boundary conditions simultaneously, we make use of the connection formula that rewrites the solutions at $x=1$ as linear combinations of solutions at $x=\infty$, 
\begin{equation}
    w_6=e^{\pi i(b-c)}\frac{\Gamma(1+c-a-b)\Gamma(b-a)}{\Gamma(1-a)\Gamma(c-a)}w_3+e^{\pi i(a-c)}\frac{\Gamma(1+c-a-b)\Gamma(a-b)}{\Gamma(1-b)\Gamma(c-b)}w_4.
\end{equation}
We shall require the coefficient in front of $w_3$ to be zero, which can be achieved by sitting on the poles of one of the $\Gamma$-functions in the denominator, 
\begin{equation}
    \frac{\Gamma\lt(1-\frac{i\tilde{\omega}}{z}\rt)\Gamma\lt(\frac{m_z}{z}\rt)}{\Gamma\lt(\frac{m_z+z-i\tilde{\omega}}{2z}\rt)^2}=0.
\end{equation}
In terms of $\omega$, we obtain the allowed QNMs as
\begin{equation}
    \omega=-ir_h^z\lt(\lt(2n+1\rt)z+m_z\rt),
\end{equation}
and reproduces the results obtained in \cite{Sybesma:2015oha}. 
We note that instead of the QNMs, one can also compute the (Fourier transformation of the) Green function (connecting the boundary and the bulk) as (refer to e.g. \cite{Skenderis:2002wp,Baggioli:2019rrs})
\begin{equation}
    G_\phi(\omega,\vec{k})=\frac{B(\omega,\vec{k})}{A(\omega,\vec{k})},
\end{equation}
where we decompose the solution that satisfies the boundary condition at the horizon as the following linear combination at the boundary $x=\infty$, 
\begin{equation}
    R(x)\sim A(\omega,\vec{k})w_3(x^z)+B(\omega,\vec{k})w_4(x^z).
\end{equation}
Then we see that the QNMs can be interpreted as some of the poles in the Green function (note that there are also additional poles in the variables e.g. $m_z$ in the Green function). This picture will be more convenient for our discussion when we take the collision limit, $r_h\rightarrow 0$. 

Another interesting case is to fix $z=2$ for general $k$. After bringing the KG equation \eqref{KG-brane} to the Schr\"odinger form, we find that the potential is proportional to
\begin{equation}
    Q(x)\propto\frac{1}{x^2\lt(x^{d+1}-1\rt)^2},
\end{equation}
and it has $d+3$ regular singularities locating at $x=0$, $x^{d+1}=1$, and $x=\infty$. The connection formulae in the second-order linear differential equation with $n$ regular singularities can be obtained from the crossing symmetry of the conformal block of the $n$-pt correlation function in 2d CFT (see e.g. \cite{Bonelli:2022ten} for $n=4$ and part of the results are presented in \cite{Lei:2023mqx}). In particular, in the case of $n=d+3=4$, the solutions are in terms of the Heun function, and its connection formulae are known. For example in \cite{Bonelli:2022ten}, the connection formulae are written down with the help of the correspondence between 4d $\cN=2$ SU(2) supersymmetric gauge theories and linear differential equations arising from the quantized Seiberg-Witten curve. We will focus on this case ($d=1$) in this article\footnote{Note that when $d=1$, there is no degree of freedom of the transverse space, $\vec{x}$. Effectively one should set $k=0$ in the equation of motion. }, in which the KG equation is solved by the Heun functions. One can build a map to put four singularities $\{0,\pm r_h,\infty\}$ to $\{0,t,1,\infty\}$ to match the Heun differential equation, 
\begin{equation}
    \lt(\frac{{\rm d}^{2}}{{\rm d}x^{2}}+\lt(\frac{\gamma}{x}+\frac{\delta}{x-1}+\frac{\epsilon}{x-t}\rt)\frac{\rm d}{{\rm d}x}+\frac{\alpha\beta x-q}{x(x-1)(x-t)}\rt)\tilde{\psi}(x)=0,\label{eq:Heun}
\end{equation}
with the constraint $\alpha+\beta=\gamma+\delta+\epsilon-1$, and compare with its Schr\"odinger form, 
\begin{equation}\label{eq:SW-eq}
 \lt(\frac{{\rm d}^2}{{\rm d}x^2}+Q_{\rm Heun}(x)\rt)\psi(x)=0,
\end{equation}
where the potential is given by 
\begin{equation}
    Q_{\rm Heun}(x)=\frac{\frac{1}{4}-a_{0}^{2}}{x^{2}}+\frac{\frac{1}{4}-a_{1}^{2}}{(x-1)^{2}}+\frac{\frac{1}{4}-a_{t}^{2}}{(x-t)^{2}}-\frac{\frac{1}{2}-a_{0}^{2}-a_{1}^{2}-a_{t}^{2}+a_{\infty}^{2}+u}{x(x-1)}+\frac{u}{x(x-t)}.\label{QSW}
\end{equation}
For the later discussion on the collision limit, we adopt the map $x=-\frac{r_h}{r}$ with $t=-1$, 
and then the dictionary is given by 
\begin{align}
    &a_0^2=1+m^2,\quad a_1^2=a_t^2=-\frac{\omega^2}{4r_h^4},\quad a_\infty^2=-\frac{\omega^2}{r_h^4},\quad u=\frac{\omega^2+(1+2m^2)r_h^4}{4r_h^4}.\label{map-brane}
\end{align}
In terms of the original parameters $\alpha$, $\beta$, $\gamma$, $\delta$, $\epsilon$, we have 
\begin{align}
    a_0^2=\frac{(1-\gamma)^2}{4},\quad  a_1^2=\frac{(1-\delta)^2}{4},\quad a_t^2=\frac{(1-\epsilon)^2}{4},\quad a_\infty^2=\frac{(\alpha-\beta)^2}{4},\label{map-a-b}
\end{align}
\begin{equation}
    u=\frac{\gamma\epsilon-2q+2\alpha\beta t-t\epsilon(\gamma+\delta)}{2(t-1)}.\label{map-u-b}
\end{equation}
and therefore 
\begin{align}
    &\gamma=1-2\sqrt{1+m^2},\quad \delta=\epsilon=\frac{i\omega}{r_h^2}+1,\quad q=0,\label{dic-Heun-1}\\
    &\alpha=1-\sqrt{1+m^2}+\frac{2i\omega}{r_h^2},\quad \beta=1-\sqrt{1+m^2}.\label{dic-Heun-2}
\end{align}
The equation \eqref{eq:Heun} can be solved as a series expansion around $x=0$, and one obtains two independent solutions as 
\begin{equation}
    \begin{aligned}
    \tilde{\psi}_{-}^{(0)}=&\text{HeunG}[t,q,\alpha,\beta,\gamma,\delta,x],\\
    \tilde{\psi}_{+}^{(0)}=&x^{1-\gamma}\text{HeunG}[t,q-(\gamma-1)(\delta t+\epsilon),\alpha- \gamma+1, \beta-\gamma+1,2-\gamma, \delta,x],
    \end{aligned}
\end{equation}
where the ${\rm HeunG}$-function is given by 
\begin{equation}
    {\rm HeunG}[t,q,\alpha,\beta,\gamma,\delta;x]=1+\frac{q}{t\gamma}x-\frac{\alpha\beta-\frac{q}{t\gamma}(\epsilon+q+\gamma+t(\gamma+\delta))}{2t(1+\gamma)}x^2+{\cal O}(x^3).
\end{equation}
Around each singularity, one can solve the differential equation to obtain two solutions in terms of the ${\rm HeunG}$-function, and different sets of solutions are connected by the connection formulae. For example, the solutions around $x=\infty$ are given by 
\begin{align}
    \tilde{\psi}^{(\infty)}_+=x^{-\alpha}{\rm HeunG}(t,q-\alpha\beta(1+t)+\alpha(\delta+t\epsilon),\alpha,\alpha-\gamma+1,\cr
    \alpha-\beta+1,\alpha+\beta+1-\gamma-\delta;t/x),
\end{align}
and 
\begin{align}
    \tilde{\psi}^{(\infty)}_-=x^{-\beta}{\rm HeunG}(t,q-\alpha\beta(1+t)+\alpha(\delta+t\epsilon),\beta,\beta-\gamma+1,\cr
    -\alpha+\beta+1,\alpha+\beta+1-\gamma-\delta;t/x).
\end{align}
The connection formulae between $x=0$ and $x=\infty$ read 
\begin{align}
    &\tilde{\psi}^{(0)}_\theta(x)=\sum_{\theta'\pm}\lt(\sum_{\sigma=\pm}\frac{e^{\frac{i\pi}{2}(\delta+\epsilon)}t^{\frac{\epsilon+\theta(1-\gamma)}{2}-\sigma a}e^{\frac{\theta}{2}\partial_{a_0}F-\frac{\theta'}{2}\partial_{a_\infty}F-\frac{\sigma}{2}\partial_{a}F}}{\Gamma\lt(1-\sigma a+\frac{\theta}{2}(1-\gamma)-\frac{\epsilon}{2}\rt)\Gamma\lt(-\sigma a+\frac{\theta}{2}(1-\gamma)+\frac{\epsilon}{2}\rt)}\rt.\cr
    &\lt.\times\frac{\Gamma(1-2\sigma a)\Gamma(-2\sigma a)\Gamma(1+\theta(1-\gamma))\Gamma(-\theta'(\alpha-\beta))}{\Gamma\lt(1-\sigma a-\frac{\theta'}{2}(\alpha-\beta)-\frac{\delta}{2}\rt)\Gamma\lt(-\sigma a-\frac{\theta'}{2}(\alpha-\beta)+\frac{\delta}{2}\rt)}\rt)\tilde{\psi}^{(\infty)}_{\theta'}(x),\label{conn-w-0inf}
\end{align}
where $F$ denotes the conformal block in the semiclassical limit and $a$ is an additional parameter that should be computed from the Matone relation \eqref{eq:Matone} (refer to Appendix \ref{a:conf-block}). The QNMs can be computed by requiring the ratio of the coefficients $A_{-+}/A_{--}$ to vanish, where we denoted the coefficients in the connection formula as 
\begin{equation}
    \tilde\psi^{(0)}_\theta=\sum_{\theta'=\pm}A_{\theta,\theta'}\tilde\psi^{(\infty)}_{\theta'}.
\end{equation}

We remark that the QNMs in the cases mentioned above are all purely imaginary, which suggests the corresponding geometry to be overdamped. This is not true for the general case of Lifshitz brane. As studied numerically in \cite{Sybesma:2015oha}, for $d\leq z+1$, all the QNMs seem to be imaginary but they become complex for $d>z+1$. It will be interesting to investigate such a transition in the future work at ODE with 7 regular singularities with similar connection formulae described above. 

The main goal of this article is to perform consistency checks that the collision limit $r_h\to 0$, in which three regular singularities collide into one irregular singularity, might be taken directly at the level of the connection formulae. 

\section{Lifshitz geometries as $r_h\rightarrow 0$ limit}\label{s:L-geometry}

When we take the limit $r_h\rightarrow 0$, the metric reduces to the form, 
\begin{equation}
    {\rm d}s^2=-r^{2z}{\rm d}t^2+\frac{1}{r^2}{\rm d}r^2+r^2{\rm d}\vec{x}^2_{d-1},
\end{equation}
which is often referred to as the Lifshitz geometry. The radial part of the massive KG equation with mass $m$ in the Lifshitz geometry is given by 
\begin{equation}
   r^{2z+2}R^{\prime\prime}(r)+r^{2z+1}(2-d+z)R'(r)+\lt(\omega^2-\lt[(d-1)z+m^2\rt]r^{2z}-k^2r^{2z-2}\rt)R(r)=0.
\end{equation}

When $k=0$, the KG equation is solved to the following two solutions, 
\begin{align}
    \bar{R}_1(r)=\lt(\frac{\omega}{2zr^{z}}\rt)^{\frac{1-d+z}{2z}}J_{-\frac{\sqrt{(d+z-1)^2+4m^2}}{2z}}\lt(\frac{\omega}{zr^z}\rt),\\
    \bar{R}_2(r)=\lt(\frac{\omega}{2zr^{z}}\rt)^{\frac{1-d+z}{2z}}J_{\frac{\sqrt{(d+z-1)^2+4m^2}}{2z}}\lt(\frac{\omega}{zr^z}\rt).
\end{align}
One can take the linear combination of these solutions to consider the modified Bessel functions $I_{\pm\mu}$, $K_\mu$ instead, where we denote 
\begin{equation}
    \mu:=\frac{\sqrt{(d+z-1)^2+4m^2}}{2z}.
\end{equation}
To require that the solution does not blow up at both $r=0$ and $r=\infty$, it should match with 
\begin{equation}
    R_{2'}(r)=\lt(\frac{\omega}{2zr^{z}}\rt)^{\frac{1-d+z}{2z}}I_{\mu}\lt(-\frac{i\omega}{zr^z}\rt),
\end{equation}
at $r=\infty$, and 
\begin{equation}
    R_3(r)=\lt(\frac{\omega}{2zr^{z}}\rt)^{\frac{1-d+z}{2z}}K_\mu\lt(\frac{i\omega}{zr^z}\rt),\label{Lg-sol-0}
\end{equation}
at $r=0$, where we assumed ${\rm Im}\omega<0$. The connection formula in this case is given by 
\begin{equation}
    K_\alpha(x)=\frac{\pi}{2}\frac{I_{-\alpha}(x)-I_\alpha(x)}{\sin(\alpha\pi)},
\end{equation}
or equivalently, 
\begin{align}
    I_\alpha(x)=\frac{e^{-\pi i\alpha}K_\alpha(x)-K_{\alpha}(e^{\pi i}x)}{\pi i},\\
    I_{-\alpha}(x)=\frac{e^{\pi i\alpha}K_\alpha(x)-K_{\alpha}(e^{\pi i}x)}{\pi i}.
\end{align}
To satisfy the desired boundary condition, we note the asymptotic behavior around $x\sim 0$, 
\begin{equation}
    I_\alpha(x)\sim \frac{1}{\Gamma(1+\alpha)}\lt(\frac{z}{2}\rt)^\alpha,
\end{equation}
the boundary condition can thus be imposed by requiring 
\begin{equation}
    \frac{1}{\Gamma(1-\mu)}=0. 
\end{equation}
In particular, when $d=z+1$, $\mu=m_z$, we obtain a set of quantized values of the mass parameters, 
\begin{equation}
    m_z=z(n+1),\quad n\in\mathbb{N}. 
\end{equation}

In the case of $z=2$, one can solve the KG equation in terms of the Whittaker functions \cite{Keeler:2013msa}. By setting $y=r^{-z}$, the KG equation reduces to 
\begin{equation}
    y^2z^2R^{\prime\prime}(y)+yz(d-1)R'(y)+\lt(y^2\omega^2+k^2y^{2/z}+z(1-d)-m^2\rt)R(y)=0.
\end{equation}
We focus on the special case of $d=1$, then for $z=2$, two solutions are found to be 
\begin{equation}
    R_5(y)=M_{\frac{ik^2}{4\omega},\frac{\sqrt{1+m^2}}{2}}(i\omega y),\quad R_6(y)=W_{\frac{ik^2}{4\omega},\frac{\sqrt{1+m^2}}{2}}(i\omega y).
\end{equation}
where we adopt the convention of the Whittaker functions $M_{\kappa,\mu}$ and $W_{\kappa,\mu}$ in terms of the confluent hypergeometric functions as 
\begin{align}
    M_{\kappa,\mu}(z):=e^{-\frac{z}{2}}z^{\mu+\frac{1}{2}}M(\mu-\kappa+\frac{1}{2},1+2\mu,z),\\
    W_{\kappa,\mu}(z):=e^{-\frac{z}{2}}z^{\mu+\frac{1}{2}}U(\mu-\kappa+\frac{1}{2},1+2\mu,z).
\end{align}
With the notation of the generalized hypergeometric function, 
\begin{equation}
    _pF_q(a_1,\dots,a_p;b_1,\dots,b_q;z):=\sum_{n=0}^\infty \frac{\prod_{i=1}^p(a_i)_n}{\prod_{j=1}^q(b_j)_n}\frac{z^n}{n!},
\end{equation}
where 
\begin{equation}
    (a)_n:=\prod_{i=1}^n(a+i-1)=\frac{\Gamma(a+n)}{\Gamma(a)},
\end{equation}
the confluent hypergeometric function $M$ can be expressed as
\begin{equation}
    M(a,c,z)=\ _1F_1(a,c,z)=\lim_{b\rightarrow \infty}\ _2F_1(a,b,c;z/b),
\end{equation}
and $U$ is defined as  
\begin{equation}
    U(a,b,z)=\frac{\Gamma(1-b)}{\Gamma(1+a-b)}M(a,b,z)+\frac{\Gamma(b-1)}{\Gamma(a)}z^{1-b}M(a+1-b,2-b,z).\label{eq:U-def}
\end{equation}
At $x=\infty$, the asymptotic behaviors of $U$ and $M$ are given by 
\begin{equation}
    U(a,b,x)\sim x^{-a},\quad M(a,b,x)\sim \Gamma(b)\lt(\frac{e^x x^{a-b}}{\Gamma(a)}+\frac{(-x)^{-a}}{\Gamma(b-a)}\rt),
\end{equation}
then to require the solution to converge at $y=\infty$, we shall pick up the solution
\begin{equation}
    R(y)=R_6(y).
\end{equation}
Since $b$ in \eqref{eq:U-def} is identified with $b\leftrightarrow 1+\sqrt{1+m^2}$, to avoid the divergence at $y=0$, we shall further require the coefficient of the term $M(a+1-b,2-b,y)$ in \eqref{eq:U-def} to vanish, i.e. 
\begin{equation}
    \frac{\Gamma(\sqrt{1+m^2})}{\Gamma\lt(\frac{1+\sqrt{1+m^2}}{2}-\frac{ik^2}{4\omega}\rt)}=0,
\end{equation}
or 
\begin{equation}
    \frac{ik^2}{4\omega}=n+\frac{1+\sqrt{1+m^2}}{2},\quad n\in\mathbb{N}.
\end{equation}

In this way, we have presented two special cases, (i) $d=z+1$, $k=0$, (ii) $z=2$, $k\neq 0$, that have known connection formulae to be solved both in the Lifshitz brane and in the $r_h\rightarrow 0$ limit. These explicit examples provide us with the perfect stage for investigating how the connection formulae behave in the collision limit $r_h\to 0$, where several regular singularities collide into an irregular one. 

\section{Collision limit of wavefunctions}\label{s:wave}

Before we discuss the collision limit of the connection formulae, let us first investigate how the wavefunctions (i.e. the solutions) behave in the limit $r_h\to 0$. There are two cases to consider explicitly in the context of the Lifshitz brane. In one case, the limit takes the hypergeometric function to the Bessel function, and in the other, it is expected to bring the Heun function to the Whittaker function.

\subsection{Case 1: hypergeometric function to Bessel function}

The connection between the hypergeometric function and the Bessel function is well-known. In terms of the generalized hypergeometric function, schematically we have in the limit $\tilde\epsilon\rightarrow 0$ (with $a$, $b$, $c$ finite), 
\begin{equation}
    _2F_1(a/\tilde\epsilon,b/\tilde\epsilon,c;\tilde\epsilon^2 \xi)\rightarrow \ _0F_1(c;ab\xi).
\end{equation}
As we further have 
\begin{equation}
    I_\alpha(x)=\frac{(x/2)^\alpha}{\Gamma(\alpha+1)}\ _0F_1\lt(\alpha+1,\frac{x^2}{4}\rt),
\end{equation}
we can relate the hypergeometric function with the modified Bessel function in the collision limit $\tilde\epsilon\rightarrow 0$ as 
\begin{equation}
    _2F_1(a/\tilde\epsilon,b/\tilde\epsilon,c;\tilde\epsilon^2 \xi)\rightarrow \Gamma(c)(\sqrt{ab\xi})^{1-c}I_{c-1}(2\sqrt{ab\xi}).\label{eq:hg-to-Bessel}
\end{equation}
In our case, we identify $\tilde\epsilon\equiv r_h^z$, $\xi\equiv r^{-2z}$. As $x^{-2z}=r_h^{2z}/r^{2z}\rightarrow 0$ vanishes and $\tilde{\omega}=\frac{\omega}{r_h^z}$ diverges in the limit $r_h\to 0$, we have
\begin{align}
    _2F_1\lt[\frac{\mp m_z+z+i\Tilde{\omega}}{2z},\frac{\mp m_z+z+i\Tilde{\omega}}{2z},1\mp \frac{m_z}{z},x^{-2z}\rt]\cr
    \rightarrow \Gamma \lt(1\mp\frac{m_z}{z}\rt)\lt(-\frac{i\omega}{2zr^z}\rt)^{\pm \frac{m_z}{z}}I_{\mp m_z/z}\lt(-\frac{i\omega}{zr^z}\rt),
\end{align}
and then in the limit 
\begin{equation}
    R_2(r)\sim r_h^{\frac{z}{2}-m_z}R_{2'}(r),
\end{equation}
where some irrelevant overall constants are omitted. 

In terms of hypergeometric functions $w_i(x)$, the above wavefunction is related to $w_4$, and a similar argument can be applied to $w_3$ to take the limit of the wavefunction. In the Lifshitz brane case, i.e. before we take the limit, two independent solutions at the horizon $r=r_h$ are characterized by $w_2(x)$ and $w_6(x)$, and we expect them to reduce to the desired wavefunctions respectively given by $K_{c-1}(2\sqrt{ab\xi})$ and $K_\alpha(2e^{\pi i}\sqrt{ab\xi})$. One can indeed check this by expanding the integral representation of the hypergeometric function around $\xi=\infty$, but equivalently we expect the connection formulae of the hypergeometric function in the collision limit reproduce the same results. This will be checked explicitly in section \ref{s:limit-hg}.

\subsection{Case 2: Heun function to Whittaker function}

As we have seen in the case of $z=2$, $d=1$, the KG equation is solved by the Heun functions. The singularities locate at $r=0$, $r=\pm r_h$ and $r=\infty$. In the limit of $r_h\rightarrow 0$, we will see that the Heun function reduces to a biconfluent Heun function with special parameters, which is further equal to a Whittaker function. 

Let us explain how the limit is realized at the level of the differential equation and the wavefunctions. We replace $z\rightarrow bz$, $t\equiv -1$, and expand the parameters to their leading orders $q\rightarrow -\frac{\tilde{q}}{b}$, $\alpha\beta\to -\frac{\alpha'}{b^2}$, $\delta\to\frac{c}{b^2}+\frac{p}{b}$, $\epsilon\to\frac{c}{b^2}+\frac{s}{b}$ in the Heun equation \eqref{eq:Heun}, to first obtain 
\begin{equation}
    \lt(\frac{d^{2}}{dz^{2}}+\lt(\frac{\gamma}{z}+\frac{\delta}{z-1/b}+\frac{\epsilon}{z+1/b}\rt)\frac{d}{dz}-\frac{\alpha' z-\tilde{q}}{b^2z(z-1/b)(z+1/b)}\rt)\tilde{\psi}(bz)=0.
\end{equation}
Then in the limit $b\to 0$, we recover the biconfluent Heun equation as 
\begin{equation}
    \lt(\frac{d^{2}}{dz^{2}}+\lt(\frac{\gamma}{z}+(s-p)-2cz\rt)\frac{d}{dz}+\frac{\alpha'z-\tilde{q}}{z}\rt)\tilde{\Psi}(z)=0,
\end{equation}
where we identified $\tilde\Psi(z)\equiv \tilde\psi(bz)$. Comparing with the standard form of the biconfluent Heun equation, 
\begin{equation}
    \frac{{\rm d}^2w}{{\rm d}z^2}+\lt[\frac{\gamma}{z}+\delta'+\epsilon' z\rt]\frac{{\rm d}w(z)}{{\rm d}z}+\frac{\alpha' z-\tilde{q}}{z}w(z)=0,
\end{equation}
we see that $\delta'=s-p$, $\epsilon'=-2c$. In our special case, $b=r_h$, $\delta'=s=p=0$, $c=i\omega$, $\tilde{q}=0$, $\epsilon'=-2i\omega$, 
\begin{equation}
    \alpha\beta=\frac{2i\omega(1-\sqrt{1+m^2})}{r_h^2}+\dots\quad \alpha'=2i\omega(\sqrt{1+m^2}-1).
\end{equation}
In this limit, the solutions to the Heun equation reduce to the biconfluent ones, 
\begin{equation}
    \tilde{\psi}^{(0)}_+(z)={\rm HeunG}[t,q,\alpha,\beta,\gamma,\delta; z]\to {\rm HeunB}[\tilde{q},\alpha',\gamma,\delta',\epsilon';z],\label{limit-HeunB1}
\end{equation}
and 
\begin{align}
    &\tilde{\psi}^{(0)}_-(z)=z^{1-\gamma}{\rm HeunG}[t,q-(\gamma-1)(t\delta+\epsilon),\alpha+1-\gamma,\beta+1-\gamma,2-\gamma,\delta;z]\cr
    &\to z^{1-\gamma}{\rm HeunB}[\tilde{q}+(\gamma-1)(s-p),\alpha'-\epsilon'(1-\gamma),2-\gamma,s-p,-2c;z],\label{limit-HeunB2}
\end{align}
where the explicit expression of the biconfluent Heun function is given by 
\begin{equation}
    {\rm HeunB}[\tilde{q},\alpha',\gamma,\delta',\epsilon';z]=1+\frac{\tilde{q}}{\gamma}z-\frac{\alpha' +\frac{\tilde{q}(\delta'-\tilde{q})}{\gamma}}{2(1+\gamma)}z^2+{\cal O}(z^3).
\end{equation}

Further in the special case $\delta'=\tilde{q}=0$, the biconfluent Heun functions are simplified to the Whittaker function as 
\begin{equation}
    {\rm HeunB}\lt[0,2(1+2k+2\mu),1+4\mu,0,2,z^{\frac{1}{2}}\rt]=e^{-z/2}z^{-\frac{1}{2}-\mu}M_{k,\mu},\label{HenB-M}
\end{equation}
and 
\begin{align}
    &z^{-2\mu}{\rm HeunB}\lt(0,2(1+2k-2\mu),1-4\mu,0,2,z^{\frac{1}{2}}\rt)\cr
    &=\frac{\Gamma(\frac{1}{2}-k+\mu)}{\Gamma(2\mu)}e^{-z/2}z^{-\frac{1}{2}-\mu}W_{k,\mu}-\frac{\Gamma(-2\mu)\Gamma(\frac{1}{2}-k+\mu)}{\Gamma(2\mu)\Gamma(\frac{1}{2}-k-\mu)}e^{-z/2}z^{-\frac{1}{2}-\mu}M_{k,\mu}.
\end{align}
This can be checked by bringing the biconfluent Heun equation to its Schr\"odinger standard form and compare with the Whittaker equation, 
\begin{equation}
    \frac{{\rm d}^2w}{{\rm d}z^2}+\lt(\frac{\frac{1}{4}-\mu^2}{z^2}+\frac{\kappa}{z}-\frac{1}{4}\rt)w=0,
\end{equation}
or compare the explicit expressions of these special functions. We remark that when $\delta'=\tilde{q}=0$, the biconfluent Heun function is given by the expansion, 
\begin{equation}
    {\rm HeunB}[0,\alpha',\gamma,0,\epsilon';z]=1-\frac{\alpha'z^2}{2(1+\gamma)}+\frac{\alpha'(\alpha'+2\epsilon')z^4}{8(1+\gamma)(3+\gamma)}+{\cal O}(z^6).
\end{equation}
It is then easy to check that 
\begin{equation}
    {\rm HeunB}[0,\alpha',\gamma,0,\epsilon';z]={\rm HeunB}[0,2\alpha'/\epsilon',\gamma,0,2;\sqrt{\epsilon'/2}z].\label{HeunB-id}
\end{equation}

An interesting but subtle point here is that around $y^{-1}:=z=\infty$, the form of the solutions to the biconfluent Heun equation changes drastically, and one should use instead the function 
\begin{align}
    &{\rm HeunB\infty}[\tilde{q},\alpha',\gamma,\delta',\epsilon';y]:=y^{\frac{\alpha'}{\epsilon'}}\lt(1-\frac{\alpha'\delta'+\tilde{q}\epsilon'}{\epsilon^{\prime 2}}y\rt.\cr
    &\lt.+\frac{\tilde{q}(\tilde{q}+\delta')\epsilon^{\prime 2}+\alpha^{\prime 2}(\delta^{\prime 2}+\epsilon')
    +\alpha'\epsilon'(2\tilde{q}\delta'+\delta^{\prime 2}+\epsilon'-\gamma' \epsilon')}{2\epsilon^{\prime 4}}y^2+{\cal O}(y^3)\rt).
\end{align}
When $\tilde{q}=\delta'=0$, it simplifies to 
\begin{align}
    &{\rm HeunB\infty}[0,\alpha',\gamma,0,\epsilon';y]:=y^{\frac{\alpha'}{\epsilon'}}\lt(1+\frac{\alpha'(\alpha'+\epsilon'(1-\gamma))}{2\epsilon^{\prime 3}}y^2\rt.\cr
    &\lt.+\frac{\alpha'(\alpha'+2\epsilon')(\alpha'+\epsilon'(1-\gamma))(\alpha'+\epsilon'(3-\gamma))}{8\epsilon^{\prime 6}}y^4+{\cal O}(y^6)\rt).
\end{align}
The other independent solution around $z=\infty$ in the special case of $\tilde{q}=\delta'=0$ is given by 
\begin{align}
    &\widetilde{\rm HeunB\infty}[0,\alpha',\gamma,0,\epsilon';y]=y^{1+\gamma-\frac{\alpha'}{\epsilon'}}\exp\lt(-\frac{\epsilon'}{2y^2}\rt)\lt(1-\frac{(\alpha'-2\epsilon')(\alpha'-(1+\gamma)\epsilon')}{2\epsilon^{\prime 3}}y^2\rt.\cr
    &\lt.+\frac{(\alpha'-2\epsilon')(\alpha'-4\epsilon')(\alpha'-(1+\gamma)\epsilon')(\alpha'-(3+\gamma)\epsilon')}{8\epsilon^{\prime 6}}y^4+{\cal O}(y^6)\rt)\cr
    &=\exp\lt(-\frac{\epsilon'}{2y^2}\rt){\rm HeunB\infty}[0,\alpha'-(1+\gamma)\epsilon',\gamma,0,-\epsilon';y].
\end{align}
We note that the identity in the last line above that relates the $\widetilde{\rm HeunB\infty}$ function with the ${\rm HeunB\infty}$ function holds for all order. One can see this by substituting $g(y)=e^{-\frac{\epsilon'}{2y^2}}h(y)$ into the biconfluent Heun equation in the $y=z^{-1}$ patch with $\delta'=\tilde{q}=0$, 
\begin{equation}
    \lt[\frac{{\rm d}^2}{{\rm d}y^2}+\lt(\frac{2-\gamma}{y}-\frac{\epsilon'}{y^3}\rt)\frac{{\rm d}}{{\rm d}y}+\frac{\alpha'}{y^4}\rt]g(y)=0,
\end{equation}
to obtain the ODE for $h(y)$ as 
\begin{equation}
    \lt[\frac{{\rm d}^2}{{\rm d}y^2}+\lt(\frac{2-\gamma}{y}+\frac{\epsilon'}{y^3}\rt)\frac{{\rm d}}{{\rm d}y}+\frac{\alpha'-(1+\gamma)\epsilon'}{y^4}\rt]h(y)=0.
\end{equation}

Around $z=\infty$, there are two independent solutions to the Heun equation given by 
\begin{align}
    z^{-\alpha}{\rm HeunG}[t,q-\alpha\beta(1+t)+\alpha(\delta+t\epsilon),\alpha,\alpha-\gamma+1,\alpha-\beta+1,\alpha+\beta+1-\gamma-\delta;t/z],
\end{align}
and 
\begin{align}
    z^{-\beta}{\rm HeunG}[t,q-\alpha\beta(1+t)+\alpha(\delta+t\epsilon),\beta,\beta-\gamma+1,-\alpha+\beta+1,\alpha+\beta+1-\gamma-\delta;t/z],
\end{align}
and by comparing the explicit expressions of series expansions, it is straightforward to see that the second HeunG function reduces to the HeunB$\infty$ function, 
\begin{align}
    \tilde\psi^{(\infty)}_-(z)=z^{-\beta}{\rm HeunG}[t,0,\beta,\beta-\gamma+1,-\alpha+\beta+1,\alpha+\beta+1-\gamma-\delta;t/(r_hz)]\cr
    \rightarrow {\rm HeunB\infty}[0,\alpha',\gamma,0,\epsilon';z^{-1}],\label{limit-HeunBinf}
\end{align}
up to a factor $r_h^{\beta}$. 

As explained previously, we expect the biconfluent Heun solutions around $z=\infty$ in the special case $\delta=\tilde{q}=0$ again reduce to the Whittaker functions, and by comparing the expansions with respect to $y=z^{-1}\sim 0$, we have 
\begin{equation}
    W_{\kappa,\mu}(y^{-1})=e^{-\frac{1}{2y}}y^{\mu-1/2}{\rm HeunB\infty}\lt[0,2(2\kappa+2\mu-1),1-4\mu,0,-2;y^{\frac{1}{2}}\rt],\label{W-HeunBinf}
\end{equation}
and 
\begin{align}
    &M_{\kappa,\mu}(y^{-1})-(-1)^{3/2+\kappa-\mu}\frac{\Gamma(1+2\mu)}{\Gamma(\frac{1}{2}+\kappa+\mu)}W_{\kappa,\mu}(y^{-1})\cr
    &=\frac{\Gamma(1+2\mu)}{\Gamma(\frac{1}{2}-\kappa+\mu)}e^{\frac{1}{2y}}y^{-\mu-1/2}{\rm HeunB\infty}\lt[0,2(2\kappa+2\mu+1),1+4\mu,0,2;y^{\frac{1}{2}}\rt].\label{MW-HeunBinf}
\end{align}

With the above information on how wavefunctions behave in the collision limit, one expects that the connection formulae of HeunG functions should match those of the Whittaker functions by taking the limit. In the next section, we examine whether this is the case.

\section{Collision limit and connection formula}\label{s:conn}

In this section, we explore the connection formulae of hypergeometric functions and those of the Heun functions obtained in \cite{Bonelli:2022ten} in the collision limit. We aim to provide some supportive evidence that the connection formulae in the collision limit still give consistent results in the limit. 

\subsection{Case 1: hypergeometric function to Bessel function}\label{s:limit-hg}

Let us begin with the case 1 of this work: $d=z+1$, $k=0$. In this case, the collision limit $r_h$ takes the hypergeometric functions to Bessel functions. It is expected that the solutions $w_2(x)$ and $w_6(x)$ reduce to the modified Bessel functions of the second kind. In the following, we take the collision limit $\tilde{\epsilon}\equiv r^z_h\to 0$ of the connection formulae of hypergeometric functions, and show that they are consistent with the connection formulae of the Bessel function, 
\begin{equation}
    K_\alpha(x)=\frac{\pi}{2}\frac{I_{-\alpha}(x)-I_\alpha(x)}{\sin(\alpha\pi)}.
\end{equation}
Again we replace $a\to a/\tilde\epsilon$ and $b\to b/\tilde\epsilon$ in the connection formulae \eqref{hg-conn-234} and \eqref{hg-conn-634}, and then we use \eqref{eq:hg-to-Bessel} and Stirling's formula, 
\begin{equation}
    \Gamma(a)=\exp\lt(-a+a\log(a)+\frac{1}{2}\log(2\pi/a)+\frac{1}{12a}+{\cal O}(a^{-3})\rt),\label{stirling-1}
\end{equation}
\begin{equation}
    \frac{\Gamma(a+b)}{\Gamma(a)}=a^b\exp\lt({\cal O}(a^{-1})\rt),\label{stirling-2}
\end{equation}
for $a$ very large and not on the negative real axis. Then we have 
\begin{align}
     w_2=&e^{-\pi ia/\tilde\epsilon}\frac{\Gamma(1+a/\tilde\epsilon+b/\tilde\epsilon-c)\Gamma(b/\tilde\epsilon-a/\tilde\epsilon)}{\Gamma(1+b/\tilde\epsilon-c)\Gamma(b/\tilde\epsilon)}w_3\cr
     &+e^{-\pi ib/\tilde\epsilon}\frac{\Gamma(1+a/\tilde\epsilon+b/\tilde\epsilon-c)\Gamma(a/\tilde\epsilon-b/\tilde\epsilon)}{\Gamma(1+a/\tilde\epsilon-c)\Gamma(a/\tilde\epsilon)}w_4\cr
    &\rightarrow\lt(\frac{r}{r_h}\rt)^{-\frac{2za}{\tilde\epsilon}}\frac{\Gamma\lt(\frac{a+b}{\tilde\epsilon}\rt)}{\Gamma\lt(\frac{b}{\tilde\epsilon}\rt)^2}\frac{\pi}{\sin(\pi d)}\lt[\lt(-\frac{i\omega}{2zr^z}\rt)^{d}I_{-d}\lt(-\frac{i\omega}{zr^z}\rt)\rt.\cr
    &\lt.-\lt(\frac{b}{\tilde\epsilon}\rt)^{2d}\lt(-\frac{i\omega}{2zr^z}\rt)^{-d}\lt(\frac{r^z}{\tilde\epsilon}\rt)^{-2d}I_{d}\lt(-\frac{i\omega}{zr^z}\rt)\rt]\cr
    &\to \lt(\frac{r}{r_h}\rt)^{-\frac{2za}{\tilde\epsilon}}\lt(\frac{i\omega}{2zr^z}\rt)^{d}\frac{\Gamma\lt(\frac{a+b}{\tilde\epsilon}\rt)}{\Gamma\lt(\frac{b}{\tilde\epsilon}\rt)^2}\frac{\pi}{\sin(\pi d)}\lt[e^{\pi id}I_{-d}\lt(-\frac{i\omega}{zr^z}\rt)-e^{-\pi id}I_{d}\lt(-\frac{i\omega}{zr^z}\rt)\rt]\cr
    &=2\lt(\frac{r}{r_h}\rt)^{-\frac{2za}{\tilde\epsilon}}\lt(\frac{i\omega}{2zr^z}\rt)^{d}\frac{\Gamma\lt(\frac{a+b}{\tilde\epsilon}\rt)}{\Gamma\lt(\frac{b}{\tilde\epsilon}\rt)^2}K_d\lt(\frac{i\omega}{zr^z}\rt),
\end{align}
where we denoted $d:=\frac{b-a}{\tilde\epsilon}=\frac{m_z}{z}$ and used $c=1$, $b\to \frac{i\omega}{2z}$. Up to some overall prefactors that can be rescaled away, it reproduces the physically allowed solution \eqref{Lg-sol-0}. We remark that when written in terms of $R(x)$ as in \eqref{R-to-w}, the factor $(x^{2z}-1)^{\frac{i\tilde{\omega}}{2z}}\sim (r/r_h)^{\frac{i\omega}{\tilde\epsilon}}$ cancels the divergent piece $(r/r_h)^{-\frac{2za}{\tilde\epsilon}}$ in $w_2$, so $R(x)$ associated to $w_2$ in the collision limit gives a well-defined wavefunction. Similarly for $w_6$, we have 
\begin{align}
		w_6=&e^{\pi i(b/\tilde\epsilon-c)}\frac{\Gamma(1+c-a/\tilde\epsilon-b/\tilde\epsilon)\Gamma(b/\tilde\epsilon-a/\tilde\epsilon)}{\Gamma(1-a/\tilde\epsilon)\Gamma(c-a/\tilde\epsilon)}w_3\cr
  &+e^{\pi i(a/\tilde\epsilon-c)}\frac{\Gamma(1+c-a/\tilde\epsilon-b/\tilde\epsilon)\Gamma(a/\tilde\epsilon-b/\tilde\epsilon)}{\Gamma(1-b/\tilde\epsilon)\Gamma(c-b/\tilde\epsilon)}w_4\cr
		&\to -e^{\frac{\pi i(a+b)}{\tilde\epsilon}}\lt(\frac{r}{r_h}\rt)^{-\frac{2za}{\tilde\epsilon}}\frac{\Gamma\lt(2-\frac{a+b}{\tilde\epsilon}\rt)}{\Gamma\lt(1-\frac{a}{\tilde\epsilon}\rt)^2}\frac{\pi}{\sin(\pi d)}\lt[\lt(-\frac{i\omega}{2zr^z}\rt)^{d}I_{-d}\lt(-\frac{i\omega}{zr^z}\rt)\rt.\cr
        &\quad \lt.-\lt(-\frac{b}{\tilde\epsilon}\rt)^{2d}\lt(-\frac{i\omega}{2zr^z}\rt)^{-d}\lt(\frac{r^z}{\tilde\epsilon}\rt)^{-2d}I_{d}\lt(-\frac{i\omega}{zr^z}\rt)\rt.\cr
        &\to -2e^{\frac{\pi i(a+b)}{\tilde\epsilon}}\lt(\frac{r}{r_h}\rt)^{-\frac{2za}{\tilde\epsilon}}\frac{\Gamma\lt(2-\frac{a+b}{\tilde\epsilon}\rt)}{\Gamma\lt(1-\frac{a}{\tilde\epsilon}\rt)^2}\lt(-\frac{i\omega}{2zr^z}\rt)^{d}K_{d}\lt(-\frac{i\omega}{zr^z}\rt).
\end{align}

\subsection{Case 2: Heun function to Whittaker function}

In the case of the reduction from the Heun function to the Whittaker function, it is desired to recover the following connection formula of the Whittaker functions, which can be obtained from \eqref{eq:U-def}, 
\begin{align}
    W_{\kappa,\mu}=\frac{\Gamma(-2\mu)}{\Gamma(-\mu-\kappa+1/2)}M_{\kappa,\mu}+\frac{\Gamma(2\mu)}{\Gamma(\mu-\kappa+1/2)}M_{\kappa,-\mu},
\end{align}
from the connection formula \eqref{conn-w-0inf} of the HeunG functions. Combining \eqref{MW-HeunBinf} and the above connection formula of the Whittaker functions, we obtain 
\begin{align}
    &(-1)^{\mu-k-3/2}\frac{\Gamma(\frac{1}{2}+k+\mu)}{\Gamma(\frac{1}{2}-k+\mu)}e^{\frac{z}{2}}z^{\mu+1/2}{\rm HeunB\infty}\lt[0,2(2k+2\mu+1),1+4\mu,0,2;z^{-\frac{1}{2}}\rt]\cr
    &=(-1)^{\mu-k-3/2}\frac{\Gamma(\frac{1}{2}+k+\mu)}{\Gamma(1+2\mu)}M_{k,\mu}(z)-\frac{\Gamma(-2\mu)}{\Gamma(-\mu-k+1/2)}M_{k,\mu}(z)\cr
    &-\frac{\Gamma(2\mu)}{\Gamma(\mu-k+1/2)}M_{k,-\mu},
\end{align}
and one can further substitute in the HeunB expression of the Whittaker function $M_{\kappa,\mu}$ \eqref{HenB-M} to have a connection formula for the HeunB functions specialized at $\tilde{q}=\delta'=0$,
\begin{align}
    &{\rm HeunB\infty}\lt[0,2(2k+2\mu+1),1+4\mu,0,2;z^{-\frac{1}{2}}\rt]\cr
    &=\frac{\Gamma(\frac{1}{2}-k+\mu)}{\Gamma(1+2\mu)}{\rm HeunB}\lt[0,2(1+2k+2\mu),1+4\mu,0,2,z^{\frac{1}{2}}\rt]\cr
    &-(-1)^{-\mu+k+3/2}\frac{\Gamma(-2\mu)\Gamma(\frac{1}{2}-k+\mu)}{\Gamma(\frac{1}{2}+k+\mu)\Gamma(-\mu-k+\frac{1}{2})}{\rm HeunB}\lt[0,2(1+2k+2\mu),1+4\mu,0,2,z^{\frac{1}{2}}\rt]\cr
    &-(-1)^{-\mu+k+3/2}\frac{\Gamma(2\mu)}{\Gamma(\frac{1}{2}+k+\mu)}z^{-2\mu}{\rm HeunB}\lt[0,2(1+2k-2\mu),1-4\mu,0,2,z^{\frac{1}{2}}\rt]\cr
    &=e^{-\pi i(\kappa+\mu+1/2)}\frac{\Gamma(-2\mu)}{\Gamma(\frac{1}{2}+\kappa -\mu)}{\rm HeunB}\lt[0,2(1+2k+2\mu),1+4\mu,0,2,z^{\frac{1}{2}}\rt]\cr
    &-e^{\pi i(-\mu+k+3/2)}\frac{\Gamma(2\mu)}{\Gamma(\frac{1}{2}+k+\mu)}z^{-2\mu}{\rm HeunB}\lt[0,2(1+2k-2\mu),1-4\mu,0,2,z^{\frac{1}{2}}\rt],\label{conn-HeunB}
\end{align}
where we used the identity 
\begin{equation}
    \Gamma(1-x)\Gamma(x)=\frac{\pi}{\sin(\pi x)}. 
\end{equation}
Rewriting the parameters in terms of $\gamma=1+4\mu$, $\alpha'=2(2\kappa+2\mu+1)$, we have 
\begin{align}
    &{\rm HeunB\infty}\lt[0,\alpha',\gamma,0,2;z^{-\frac{1}{2}}\rt]=e^{-\pi i(\kappa+\mu+1/2)}\frac{\Gamma\lt(\frac{1-\gamma}{2}\rt)}{\Gamma\lt(\frac{\alpha'}{4}+\frac{1-\gamma}{2}\rt)}{\rm HeunB}\lt[0,\alpha',\gamma,0,2,z^{\frac{1}{2}}\rt]\cr
    &-e^{\pi i(-\mu+k+3/2)}\frac{\Gamma\lt(\frac{\gamma-1}{2}\rt)}{\Gamma\lt(\frac{\alpha'}{4}\rt)}z^{\frac{1-\gamma}{2}}{\rm HeunB}\lt[0,\alpha'+2-2\gamma,2-\gamma,0,2,z^{\frac{1}{2}}\rt].\label{conn-HeunB-ga1}
\end{align}
Using \eqref{HeunB-id}, we have $\alpha'=1+\gamma$ when we set $\epsilon'=2$. Then, in this case, the connection formula is further simplified to  
\begin{align}
    &{\rm HeunB\infty}\lt[0,\alpha',\gamma,0,2;z^{-\frac{1}{2}}\rt]=\frac{e^{-\frac{\pi i(1+\gamma)}{4}}2^{-\frac{1+\gamma}{2}}}{\sqrt{\pi}}\lt(\Gamma\lt(\frac{1-\gamma}{4}\rt){\rm HeunB}\lt[0,\alpha',\gamma,0,2,z^{\frac{1}{2}}\rt]\rt.\cr
    &\lt.+\Gamma\lt(-\frac{1-\gamma}{4}\rt)z^{\frac{1-\gamma}{2}}{\rm HeunB}\lt[0,\alpha'+2-2\gamma,2-\gamma,0,2,z^{\frac{1}{2}}\rt]\rt),\label{conn-HeunB-ga}
\end{align}
where we used the Legendre duplication formula, 
\begin{equation}
    \Gamma(2z)=\frac{2^{2z-1}\Gamma(z)\Gamma(z+1/2)}{\sqrt{\pi}}.
\end{equation}
We remark that when $\kappa=0$, one can also obtain the above formula directly from \eqref{W-HeunBinf}.

Schematically we can write the connection formula of the Heun function, \eqref{conn-w-0inf}, as 
\begin{equation}
    \tilde\psi^{(0)}_\theta=\sum_{\theta'=\pm}A_{\theta,\theta'}\tilde\psi^{(\infty)}_{\theta'},
\end{equation}
then we have 
\begin{equation}
    \tilde\psi^{(\infty)}_-=\frac{1}{A_{+-}A_{-+}-A_{--}A_{++}}\lt(A_{-+}\tilde\psi^{(0)}_+-A_{++}\tilde\psi^{(0)}_-\rt).\label{conn-w-inf0}
\end{equation}
From the collision limit of the wavefunctions \eqref{limit-HeunB1}, \eqref{limit-HeunB2} and \eqref{limit-HeunBinf}, we see that the above equation \eqref{conn-w-inf0} takes a similar form as the connection formula \eqref{conn-HeunB} deduced from the Whittaker functions, by replacing $z\to z^{\frac{1}{2}}$. 
We naively expect that if dealt properly, the connection formulae of the Heun functions will reduce to those of the biconfluent Heun functions in the collision limit $r_h\to 0$, and therefore \eqref{conn-w-inf0} is expected to reduce to \eqref{conn-HeunB}. 

\begin{figure}
    \centering
    \includegraphics[width=8cm]{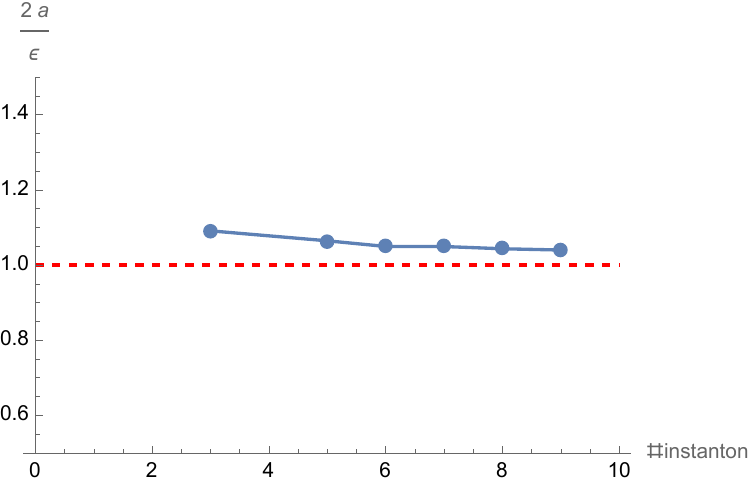}
    \caption{The plot of $\frac{2a}{\epsilon}$ against instanton numbers for $m=0.53$, $\omega=0.2-0.13i$, $r_h=0.0005$. }
    \label{fig:num-test-a}
\end{figure}

This statement, unfortunately, is highly non-trivial, as one can see in the connection formula \eqref{conn-w-0inf} and the dictionary \eqref{dic-Heun-1} and \eqref{dic-Heun-2}, almost all the arguments in the $\Gamma$-function prefactors become divergent at the order of $r_h^{-2}$ in the collision limit, and in the connection formula we wish to recover, \eqref{conn-HeunB},  there are a bunch of $\Gamma$-functions with finite arguments. One of the necessary conditions to recover the connection formula \eqref{conn-HeunB} from that of the Heun functions \eqref{conn-w-0inf} is thus to have the divergence in part of the arguments of the $\Gamma$-functions to cancel, and the most naive one is to require 
\begin{equation}
    a=\frac{\epsilon}{2}+{\cal O}(r_h^0),\label{nec-cond}
\end{equation}
in the connection formula of the Heun functions. From \eqref{dic-Heun-1}, we see that $\epsilon$ diverges as ${\cal O}(r_h^{-2})$, and it is also easy to check that the first several expansion coefficients of $a$ found from the Matone relation \eqref{eq:Matone} and \eqref{Matone-u-a} are also of the order ${\cal O}(r_h^{-2})$. We first performed numerical studies on the brute force evaluation of $a$ from the Matone relation \eqref{eq:Matone}. We found that the ratio $\frac{2a}{\epsilon}$ at small $r_h$ gradually approaches the expected value $1$ when increasing the number of instanton corrections in the numerical computation (see Figure \ref{fig:num-test-a}). Since we need to set $t=-1$ in the computation of $a$, we note that there is no guarantee of the convergence in the instanton-expansion of $a$. However, in the cases we studied, the numerical results show a tendency to converge. Moreover, we also evaluated $a$ in an indirect way, the monodromy approach developed in \cite{Zamolodchikov:426555,Harlow:2011ny}, in which we expand $a$ with respect to $r_h$ instead of the instanton parameter $t$. The leading order calculation shows that the necessary condition \eqref{nec-cond} indeed holds (see Appendix \ref{a:monodromy} for more detailed discussions). This provides a piece of supportive evidence that one may take the collision limit directly to derive the connection formulae for special functions of confluent types. We further analyzed the leading-order results of the coefficients $A_{\pm +}$ in the Appendix \ref{a:Heun} and discussed some additional necessary conditions to be satisfied to fully reproduce the expected connection formula \eqref{conn-HeunB-ga}. 

\paragraph{Remark:} We only presented the numerical results in Figure \ref{fig:num-test-a} up to 9-instanton level. Due to the exponential increasing of number of terms in the instanton expansion, it is exponentially more costing to compute the higher-order corrections with the improvement slowing down very quickly. The gauge-theory approach is therefore desired to be simplified by finding some analytic pattern in the semiclassical limit of the conformal block computation to obtain high precision numerical results.

\section{Conclusion and discussions}

In this article, we performed case studies on the collision behavior $r_h\to 0$ of the connection formulae from the Lifshitz brane to the Lifshitz geometry. In particular, we considered two special cases: (i) $d=z+1$ (with $k=0$); (ii) $d=1$, $z=2$, where in both cases the collision limit takes ODEs with only regular singularities to those with one irregular singularity. In case (i), the collision limit takes the hypergeometric function to the Bessel functions, while in case (ii), the Heun functions reduce to the biconfluent Heun functions, and are further expected to be expressed in terms of the Whittaker functions in the special case of Lifshitz geometry. We tested the naive idea of whether one can directly take the collision limit at the level of the connection formulae. In case (i), it works in a clear way. In case (ii), due to the complexity of the computation, we still cannot fully check it, but we provided a piece of supportive evidence by examining a key necessary condition for the idea of collision limit to work. 

In general, the collision limit considered in this article bringing three regular singularities into one irregular singularity takes a Heun function into a biconfluent Heun function, and the connection formulae for the latter is unkown so far. In this article, we focused on the reproduction of the known connection formulae in the special case in which the biconfluent Heun function degenerates to the Whittaker function. It is more desired in the future works to derive the generic connection formulae of the biconfluent Heun function by dealing the collision limit properly. The biggest obstruction at the current stage is that it is still difficult to manipulate the conformal block or equivalently the instanton partition function and its derivatives in the collision limit. 

In terms of supersymmetric gauge theories, the biconfluent Heun function explored in this work is related to the AD theory labeled by $(A_1,D_4)$. Although there have been some efforts to develop techniques \cite{Choi:2015idw,Rim:2016msx,Nishinaka:2019nuy,Kimura:2020krd,Kimura:2022yua,Fucito:2023plp,Fucito:2023txg,Hamachika:2024efr} to compute the AD partition function order-by-order, we still do not have an efficient method to compute the instanton partition function of AD theories. In our approach, we will further need to check the conditions \eqref{cond-1} and \eqref{cond-2}, and it is desired to use the AD partition function to check them directly as a future work. Moreover by comparing \eqref{conn-HeunB-ga} and the finite $\Gamma$-function part in \eqref{Amp}, \eqref{App}, we see that the instanton partition function in the collision limit may further contain terms of the form $\log\Gamma(c_1+c_2\gamma)$ for some c-numbers $c_{1,2}$. This reminds us of the perturbative part of the Nekrasov partition function, e.g. the perturbative contribution of the matter hypermultiplets is given by 
\begin{equation}
    \lim_{\epsilon_2\rightarrow 0}\epsilon_2\partial_a F^{\rm pert}_{\rm fund.}=\sum_{f=1}^{N_f}\lt[\frac{2a}{\hbar}\log\frac{t}{\hbar}+\log\frac{\Gamma\lt(1+\frac{a-m_f}{\hbar}\rt)}{\Gamma\lt(1-\frac{a+m_f}{\hbar}\rt)}\rt].
\end{equation}
This suggests that in the Nekrasov-Shatashvili limit, some perturbative contributions might be canceled by the instanton part in the collision limit where a Lagrangian theory flows into an AD theory. We plan to explore such properties and the connection formulae associated to the AD theories from the aspect of direct computations of the partition function in our follow-up work. 

Last but not least, we comment that the collision limit in the context of blackholes often means the coincidence of two horizons, e.g. the inner and outer horizons, i.e. the Cauchy and event horizons of Reissner–Nordstr\"om blackholes. This corresponds to the extremal limit of charged blackholes. Once our method of taking the collision limit of the connection formulae gets established in the future work, it will help to understand the behavior of QNMs in the extremal limit.

\paragraph{Acknowledgement}

We would like to thank Daniele Gregori, Yang Lei, Hongfei Shu, Jie-qiang Wu, Kilar Zhang, Hao Zou for inspiring discussions on relevant topics. R.Z. is supported by National Natural Science Foundation of China No. 12105198 and the High-level personnel project of Jiangsu Province (JSSCBS20210709).

\appendix

\section{Conformal block and Matone relation}\label{a:conf-block}

In this Appendix, we review some necessary information about the conformal block (in the semiclassical limit) for the calculation of the connection formulae. 

Consider a four-point conformal block in the 2d CFT with central charge $c=1+6Q^2$ with $Q=b+b^{-1}$. The conformal dimensions of four primary fields are taken to be $\Delta_i=\frac{Q^2}{4}-\alpha_i^2$ for $i=0,1,t,\infty$ standing for the positions of the primary operators. The semiclassical conformal block $F$ is defined as the log of the conformal block in the semiclassical limit $b\to 0$, 
\begin{align}
    F:=\lim_{b\rightarrow 0}b^2\log\mathfrak{F}(\Delta_\infty,\Delta_1,\Delta,\Delta_t,\Delta_0;t),\label{def:F}
\end{align}
where in practical one can compute the conformal block from the instanton partition function of 4d ${\cal N}=2$ supersymmetric gauge theories with the Alday-Gaiotto-Tachikawa relation \cite{Alday:2009aq}, 
\begin{align}
 &\mathfrak{F}(\Delta_\infty,\Delta_1,\Delta,\Delta_t,\Delta_0;t)=t^{\Delta-\Delta_t-\Delta_0}(1-t)^{-2(\a_t+\frac{Q}{2})(\frac{Q}{2}-\a_1)}\cr
    &\times Z^{inst}_{SU(2)\ N_f=4}(\alpha,\{2\epsilon_+-m_1,2\epsilon_+-m_2,m_3,m_4\};t),\label{eq:AGT}
\end{align}
and $\Delta=\frac{Q^2}{4}-\alpha^2$. The instanton partition function is given by the explicit form, 
\begin{equation}
    Z^{\rm inst}_{SU(2)\ N_f=4}(a,\{m_f\}_{f=1}^4;\mathfrak{q}):=\sum_{\mu,\nu}\mathfrak{q}^{|\mu|+|\nu|}Z^{\rm vect}_{\mu\nu}(a)Z^{\rm fund}_{\mu\nu}(a,\{m_f\}_{f=1}^4),
\end{equation}
where the summation over$\mu$ and $\nu$ runs over all Young diagrams, and each building block of the instanton partition function is written in terms of the Nekrasov factor, 
\begin{align}
    N_{\lambda\nu}(a,\epsilon_1,\epsilon_2):=\prod_{(i,j)\in\lambda}\lt(a+\epsilon_1(-\nu^t_j+i)+\epsilon_2(\lambda_i-j+1)\rt)\cr
    \times \prod_{(i,j)\in\nu}\lt(a+\epsilon_1(\lambda^t_j-i+1)+\epsilon_2(-\nu_i+j)\rt),
\end{align}
as 
\begin{equation}
    Z^{\rm vect}_{\mu\nu}(a):=N^{-1}_{\mu\mu}(1,\epsilon_1,\epsilon_2)N^{-1}_{\mu\nu}(2a,\epsilon_1,\epsilon_2)N^{-1}_{\nu\nu}(1,\epsilon_1,\epsilon_2)N^{-1}_{\nu\mu}(-2a,\epsilon_1,\epsilon_2),
\end{equation}
and 
\begin{align}
    Z^{\rm fund}_{\mu\nu}(a,\{m_f\}_{f=1}^{N_f}):=\prod_{f=1}^{N_f}N_{\mu\emptyset}(a-m_f,\epsilon_1,\epsilon_2)N_{\nu\emptyset}(-a-m_f,\epsilon_1,\epsilon_2).
\end{align}
The mass parameters $m_f$ are related to $\alpha_i$'s as 
\begin{align}
    & m_1=\alpha_1+\alpha_2+\frac{Q}{2},\quad m_2=-\alpha_1+\alpha_2+\frac{Q}{2},\\
    & m_3=\alpha_3+\alpha_4+\frac{Q}{2},\quad m_4=\alpha_3-\alpha_4+\frac{Q}{2}.
\end{align}

The four-point conformal block connects to the Heun equation \eqref{eq:Heun} in the semiclassical limit $b\to 0$, and thus the parameters in the Heun equation can be written in terms of the combinations $a_i:=\lim_{b\to 0}\alpha_i/b$ and $a:=\lim_{b\to 0}\alpha/b$ kept finite in the limit as 
\begin{align}
    a_0^2=\frac{(1-\gamma)^2}{4},\quad  a_1^2=\frac{(1-\delta)^2}{4},\quad a_t^2=\frac{(\alpha+\beta-\gamma-\delta)^2}{4},\quad a_\infty^2=\frac{(\alpha-\beta)^2}{4},\label{dic-Heun-BPZ-1}
\end{align}
and 
\begin{equation}
    u=\frac{\gamma\epsilon-2q+2\alpha\beta t-t\epsilon(\gamma+\delta)}{2(t-1)},\label{dic-Heun-BPZ-2}
\end{equation}
where $u$ is defined through the Matone relation \cite{Matone:1995rx}, 
\begin{equation}
    u:=\lim_{b\rightarrow 0}b^2t\partial_t\log\mathfrak{F}(\a_\infty,\a_1,\a,\a_t,\a_0;t).\label{eq:Matone}
\end{equation}
We note that the parameter $a$ does not appear directly in the differential equation \eqref{eq:Heun}, and one needs to solve the Matone relation inversely to find the expression of $a$ in terms of the parameters in the Heun equation. More explicitly, we have 
\begin{align}
    u=&-a^2+\frac{-2(\bar{m}_1+\bar{m}_2)+2(\bar{m}_1^2+\bar{m}_2^2)}{4}+(\bar{m}_1+\bar{m}_2)\lt(1-\frac{\bar{m}_3+\bar{m}_4}{2}\rt)\frac{t}{1-t}\cr
    &+\lim_{b\rightarrow 0}b^2t\partial_{t}\log  Z^{inst}_{SU(2)\ N_f=4}(a,\{2\epsilon_+-m_1,2\epsilon_+-m_2,m_3,m_4\};t).
\end{align}
One can solve $a^2$ inversely as a series expansion over $t$, and obtain one of the branches as 
\begin{align}
    a=&\frac{1}{2}\sqrt{-1+4a_0^2+4a_t^2-4u}\cr
    &-\frac{(-1+2a_0^2+2a_1^2-2a_\infty^2+2a_t^2-2u)(-1+4a_t^2-2u)}{4(-1+2a_0^2+2a_t^2-2u)\sqrt{-1+4a_0^2+4a_t^2-4u}}t+{\cal O}(t^2).\label{Matone-u-a}
\end{align}
We can then substitute \eqref{dic-Heun-1} and \eqref{dic-Heun-2} into the above expression to construct the dictionary for $a$. Unfortunately, no closed-form formula is known for \eqref{Matone-u-a}.  

\section{Connection formulae of hypergeometric functions}

Let us summarize the connection formulae of the hypergeometric functions in this Appendix.  
\begin{align}
    w_3=\frac{\Gamma(1+a-b)\Gamma(1-c)}{\Gamma(1-b)\Gamma(1+a-c)}w_1+e^{\pi i(c-1)}\frac{\Gamma(1+a-b)\Gamma(c-1)}{\Gamma(c-b)\Gamma(a)}w_5,\\
    w_4=\frac{\Gamma(1+b-a)\Gamma(1-c)}{\Gamma(1-a)\Gamma(1+b-c)}w_1+e^{\pi i(c-1)}\frac{\Gamma(1+b-a)\Gamma(c-1)}{\Gamma(c-a)\Gamma(b)}w_5,
\end{align}
\begin{align}
    w_2=\frac{\Gamma(1+a+b-c)\Gamma(1-c)}{\Gamma(1+a-c)\Gamma(1+b-c)}w_1+\frac{\Gamma(1+a+b-c)\Gamma(c-1)}{\Gamma(a)\Gamma(b)}w_5,\\
    w_6=\frac{\Gamma(1+c-a-b)\Gamma(1-c)}{\Gamma(1-a)\Gamma(1-b)}w_1+\frac{\Gamma(1+c-a-b)\Gamma(c-1)}{\Gamma(c-a)\Gamma(c-b)}w_5,
\end{align}
\begin{align}
    w_2=e^{-\pi ia}\frac{\Gamma(1+a+b-c)\Gamma(b-a)}{\Gamma(1+b-c)\Gamma(b)}w_3+e^{-\pi ib}\frac{\Gamma(1+a+b-c)\Gamma(a-b)}{\Gamma(1+a-c)\Gamma(a)}w_4,\label{hg-conn-234}\\
    w_6=e^{\pi i(b-c)}\frac{\Gamma(1+c-a-b)\Gamma(b-a)}{\Gamma(1-a)\Gamma(c-a)}w_3+e^{\pi i(a-c)}\frac{\Gamma(1+c-a-b)\Gamma(a-b)}{\Gamma(1-b)\Gamma(c-b)}w_4.\label{hg-conn-634}
\end{align}
where
\begin{align}
    &w_1(r)={}_2F_1[a,b,c,r],\quad w_5(r)=z^{1-c}\ _2F_1(1+a-c,1+b-c,2-c,r),\\
    &w_3(x)=(-x)^{-a}\ _2F_1[a,1+a-c,1+a-b,x^{-1}],\\
    &w_4(x)=(-x)^{-b}\ _2F_1[b,1+b-c,1+b-a,x^{-1}],\\
&w_2(x)=\ _2F_1[a,b,1+a+b-c,1-x],\\
    &w_6(x)=(1-x)^{c-a-b}\ _2F_1[c-a,c-b,1+c-a-b,1-x].
\end{align}

\section{Connection formulae of Heun functions and collision limit}\label{a:Heun}

In this Appendix, we first summarize the connection formulae proposed in \cite{Bonelli:2022ten}, and then investigate the behavior of the connection formula between $z=0$ and $z=\infty$ in the collision limit $r_h\to 0$. Three fundamental connection formulae are known as 
\begin{align}
    &\tilde\psi^{(0)}_\theta(z)=\sum_{\theta'=\pm}(1-t)^{-\frac{\delta}{2}}t^{(1+\theta)\frac{1-\gamma}{2}-(1+\theta')\frac{1-\epsilon}{2}}e^{\frac{\theta}{2}\partial_{a_0}F-\frac{\theta'}{2}\partial_{a_t}F}\cr
    &\times\frac{\Gamma\lt(\theta'(\epsilon-1)\rt)\Gamma\lt(1+\theta(1-\gamma)\rt)}{\Gamma\lt(\frac{1}{2}-a+\frac{\theta}{2}(1-\gamma)-\frac{\theta'}{2}(1-\epsilon)\rt)\Gamma\lt(\frac{1}{2}+a+\frac{\theta}{2}(1-\gamma)-\frac{\theta'}{2}(1-\epsilon)\rt)}\tilde\psi^{(t)}_{\theta'}(z),\label{conn-w-0t}\\
    &\tilde{\psi}^{(0)}_\theta(z)=\sum_{\theta'\pm}\lt(\sum_{\sigma=\pm}\frac{e^{\frac{i\pi}{2}(\delta+\epsilon)}t^{\frac{\epsilon+\theta(1-\gamma)}{2}-\sigma a}e^{\frac{\theta}{2}\partial_{a_0}F-\frac{\theta'}{2}\partial_{a_\infty}F-\frac{\sigma}{2}\partial_{a}F}}{\Gamma\lt(1-\sigma a+\frac{\theta}{2}(1-\gamma)-\frac{\epsilon}{2}\rt)\Gamma\lt(-\sigma a+\frac{\theta}{2}(1-\gamma)+\frac{\epsilon}{2}\rt)}\rt.\cr
    &\lt.\times\frac{\Gamma(1-2\sigma a)\Gamma(-2\sigma a)\Gamma(1+\theta(1-\gamma))\Gamma(-\theta'(\alpha-\beta))}{\Gamma\lt(1-\sigma a-\frac{\theta'}{2}(\alpha-\beta)-\frac{\delta}{2}\rt)\Gamma\lt(-\sigma a-\frac{\theta'}{2}(\alpha-\beta)+\frac{\delta}{2}\rt)}\rt)\tilde{\psi}^{(\infty)}_{\theta'}(z),\\
    &\tilde\psi^{(1)}_{\theta}(z)=-(1-t)^{\frac{\epsilon}{2}}\sum_{\theta'=\pm}\frac{\Gamma\lt(1+\theta(1-\delta)\rt)\Gamma\lt(\theta'(\beta-\alpha)\rt)e^{\frac{\theta}{2}\partial_{a_1}F-\frac{\theta'}{2}\partial_{a_\infty}F}}{\prod_{\sigma=\pm}\Gamma\lt(\frac{1}{2}+\sigma a+\frac{\theta}{2}(1-\delta)+\frac{\theta'}{2}(\beta-\alpha)\rt)}\tilde\psi^{(\infty)}_{\theta'}(z),\label{conn-w-1inf}
\end{align}
and connection formulae of other types can be derived from them. 
In this article, expressing $\tilde\psi^{(\infty)}_\pm$ in terms of $\tilde\psi^{(0)}_\pm$ becomes essential, and as discussed in \eqref{conn-w-inf0}, by denoting the decomposition coefficients of $\tilde{\psi}^{(0)}_\theta$ in terms of $\tilde{\psi}^{(\infty)}_{\theta'}$ as $A_{\theta\theta'}$, the inverse connection formula is given by 
\begin{equation}
    \tilde\psi^{(\infty)}_-=\frac{A_{-+}}{A_{+-}A_{-+}-A_{--}A_{++}}\tilde\psi^{(0)}_+-\frac{A_{++}}{A_{+-}A_{-+}-A_{--}A_{++}}\tilde\psi^{(0)}_-.
\end{equation}
Let us first evaluate $A_{+-}A_{-+}-A_{--}A_{++}$, and it turned out to be 
\begin{align}
    A_{+-}A_{-+}-A_{--}A_{++}=\frac{1-\gamma}{\alpha+\frac{1-\gamma}{2}}\frac{\cos\lt(\frac{\pi}{2}(-2a+\alpha+\frac{1-\gamma}{2}-\delta)\rt)}{\cos\lt(\frac{\pi}{2}(2a-\alpha-\frac{1-\gamma}{2}-\delta)\rt)}.
\end{align}
Using the dictionary \eqref{dic-Heun-1} and \eqref{dic-Heun-2}, especially $\alpha=2\epsilon+\frac{\gamma}{2}-\frac{3}{2}$, we first obtain 
\begin{align}
    \cos\lt(\frac{\pi}{2}(-2a+\alpha+\frac{1-\gamma}{2}-\delta)\rt)=\frac{\pi}{\Gamma\lt(-d_0\rt)\Gamma(1+d_0)},
\end{align}
and thus 
\begin{equation}
    A_{+-}A_{-+}-A_{--}A_{++}\to \frac{\pi(1-\gamma)e^{-\pi i\epsilon}}{\epsilon \Gamma\lt(-d_0\rt)\Gamma(1+d_0)},
\end{equation}
where we assumed the real part of $\omega$ to be positive, and one can simply change the sign on the shoulder of exp in the opposite case. 
We denoted $a=\frac{\epsilon}{2}+d_0$ following \eqref{nec-cond}, and $\alpha-\beta\sim 2(\epsilon-1)+\nu$ (in our case $\nu=0$).
The coefficients $A_{-+}$ and $A_{++}$ are also simplified in the collision limit as 
\begin{align}
    A_{-+}\to& \frac{\Gamma(1-\epsilon-2d_0)\Gamma(-\epsilon-2d_0)\Gamma(\gamma)\Gamma(2-2\epsilon-\nu)(e^{\pi i})^{\epsilon+\frac{1}{2}(1-\gamma)}e^{-\frac{1}{2}\partial_{a_0}F-\frac{1}{2}\partial_{a_\infty}F-\frac{1}{2}\partial_aF}}{\Gamma\lt(\frac{1}{2}(1+\gamma)-\epsilon-d_0\rt)\Gamma(-\frac{1}{2}(1-\gamma)-d_0)\Gamma(2-2\epsilon-\nu/2-d_0)\Gamma(1-\epsilon-\nu/2-d_0)}\cr
    &+\frac{\Gamma(1+\epsilon+2d_0)\Gamma(\epsilon+2d_0)\Gamma(\gamma)\Gamma(2-2\epsilon-\nu)(e^{\pi i})^{\frac{1}{2}(1-\gamma)}e^{-\frac{1}{2}\partial_{a_0}F-\frac{1}{2}\partial_{a_\infty}F+\frac{1}{2}\partial_aF}}{\Gamma\lt(\frac{1}{2}(1+\gamma)+d_0\rt)\Gamma(\epsilon-\frac{1}{2}(1-\gamma)+d_0)\Gamma(2-\epsilon-\nu/2+d_0)\Gamma(1-\nu/2+d_0)},\cr
    \sim& \frac{\Gamma(1-\epsilon-2d_0)\Gamma(-\epsilon-2d_0)\Gamma(\gamma)\Gamma(2-2\epsilon-\nu)(e^{\pi i})^{\epsilon+\frac{1}{2}(1-\gamma)}e^{-\frac{1}{2}\partial_{a_0}F-\frac{1}{2}\partial_{a_\infty}F-\frac{1}{2}\partial_aF}}{\Gamma\lt(\frac{1}{2}(1+\gamma)-\epsilon-d_0\rt)\Gamma(-\frac{1}{2}(1-\gamma)-d_0)\Gamma(2-2\epsilon-\nu/2-d_0)\Gamma(1-\epsilon-\nu/2-d_0)}\cr
    \to& (-\epsilon)^{-d_0-\frac{1+\gamma}{2}}2^{d_0-\frac{\nu}{2}}\frac{\Gamma(\gamma)}{\Gamma\lt(\frac{\gamma-1}{2}-d_0\rt)}e^{-\frac{1}{2}\partial_{a_0}F-\frac{1}{2}\partial_{a_\infty}F-\frac{1}{2}\partial_aF},\label{Amp}
\end{align}
where we used \eqref{dic-Heun-1}, \eqref{dic-Heun-2} and Stirling's formula \eqref{stirling-1} to see that the first term dominates over the second. Similarly we have 
\begin{align}
    A_{++}\to& \frac{\Gamma(1-\epsilon-2d_0)\Gamma(-\epsilon-2d_0)\Gamma(2-\gamma)\Gamma(2-2\epsilon-\nu)(e^{\pi i})^{\epsilon-\frac{1}{2}(1-\gamma)}e^{\frac{1}{2}\partial_{a_0}F-\frac{1}{2}\partial_{a_\infty}F-\frac{1}{2}\partial_aF}}{\Gamma\lt(-\frac{1}{2}(1+\gamma)-\epsilon-d_0\rt)\Gamma(\frac{1}{2}(1-\gamma)-d_0)\Gamma(2-2\epsilon-\nu/2-d_0)\Gamma(1-\epsilon-\nu/2-d_0)}\cr
    &+\frac{\Gamma(1+\epsilon+2d_0)\Gamma(\epsilon+2d_0)\Gamma(2-\gamma)\Gamma(2-2\epsilon-\nu)(e^{\pi i})^{-\frac{1}{2}(1-\gamma)}e^{\frac{1}{2}\partial_{a_0}F-\frac{1}{2}\partial_{a_\infty}F+\frac{1}{2}\partial_aF}}{\Gamma\lt(d_0-\frac{1}{2}(1+\gamma)\rt)\Gamma(d_0+\epsilon+\frac{1}{2}(1-\gamma))\Gamma(d_0+2-\epsilon-\nu/2)\Gamma(d_0+1-\nu/2)}\cr
    \sim& \frac{\Gamma(1-\epsilon-2d_0)\Gamma(-\epsilon-2d_0)\Gamma(2-\gamma)\Gamma(2-2\epsilon-\nu)(e^{\pi i})^{\epsilon-\frac{1}{2}(1-\gamma)}e^{\frac{1}{2}\partial_{a_0}F-\frac{1}{2}\partial_{a_\infty}F-\frac{1}{2}\partial_aF}}{\Gamma\lt(-\frac{1}{2}(1+\gamma)-\epsilon-d_0\rt)\Gamma(\frac{1}{2}(1-\gamma)-d_0)\Gamma(2-2\epsilon-\nu/2-d_0)\Gamma(1-\epsilon-\nu/2-d_0)}\cr
    \to &(-\epsilon)^{-d_0+\frac{1+\gamma}{2}}2^{d_0-\frac{\nu}{2}}\frac{\Gamma(2-\gamma)}{\Gamma\lt(\frac{1-\gamma}{2}-d_0\rt)}e^{\frac{1}{2}\partial_{a_0}F-\frac{1}{2}\partial_{a_\infty}F-\frac{1}{2}\partial_aF}.\label{App}
\end{align}
Unfortunately we still do not have a closed form of $F$ in the collision limit, and it becomes very difficult to go further to compare with the expected connection formula for the HeunB function \eqref{conn-HeunB-ga}. Yet we can still extract out some expected properties for the derivatives of $F$ in the collision limit. Note that $\epsilon={\cal O}(r_h^2)$, and $\tilde\psi^{(0)}_-$ differs from $\tilde\psi^{(0)}_+$ by an overall factor $z^{1-\gamma}$, which is further rescaled by a factor $r_h^{\gamma-1}$, it is then expected that 
\begin{equation}
    e^{\partial_{a_0}F}\to \epsilon^{-\frac{1-\gamma}{2}-2}.\label{cond-1}
\end{equation}
Similarly $\tilde\psi^{(\infty)}_-$ holds an overall factor $z^{-\beta}$, and from \eqref{limit-HeunBinf} it is necessary to have $r_h^\beta\sim \epsilon^{-\frac{1+\gamma}{4}}$ from the r.h.s. of the connection formula. This suggests 
\begin{equation}
    e^{-\frac{1}{2}\partial_{a_\infty}F-\frac{1}{2}\partial_aF}\to e^{-2\pi i\epsilon}\epsilon^{\frac{\gamma}{2}-2+d_0}.\label{cond-2}
\end{equation}

\section{Monodromy approach to semiclassical conformal blocks}\label{a:monodromy}

In this Appendix, we first review the monodromy approach to compute the conformal blocks, and then we apply it in the collision limit $r_h\to 0$ of the semiclassical conformal block. 

Let us denote $\delta_i=b^2\Delta_i$ and $\delta=b^2\Delta$, then the BPZ equation of the 4-pt correlation function with a degenerate field inserted is given by 
\begin{equation}
    \lt[\frac{{\rm d}^2}{{\rm d}z^2}+\frac{\delta_0}{z^2}+\frac{\delta_t}{(z-t)^2}+\frac{\delta_1}{(z-1)^2}-\frac{\delta_0+\delta_t+\delta_1-\delta_\infty}{z(z-1)}+\frac{C_2t(1-t)}{z(z-t)(z-1)}\rt]\psi(z)=0,\label{eq:BPZ}
\end{equation}
with 
\begin{equation}
    C_2=\partial_t{\cal F}_{cl},\quad \mathfrak{F}\sim e^{-\frac{c}{6}{\cal F}_{cl}}\sim e^{-b^{-2}{\cal F}_{cl}},
\end{equation}
in the semiclassical limit $c\to\infty$. The above BPZ equation reduces to the Heun equation in the semiclassical limit, and that is why the semiclassical conformal block appears in the connection formulae of the Heun functions. When the conformal weight $\Delta$ is much larger than $c$, i.e. $\Delta/c\gg 1$, the asymptotic behavior of the conformal block fixed by ${\cal F}_{cl}$ can be found from the monodromy method developed in \cite{Zamolodchikov:426555,Harlow:2011ny}. Let us first briefly review the logic behind the calculation, and then apply it to our case further with additional $\Delta_i\gg c$. 

The monodromy property of two solutions $\psi_\pm(z)$ circling around both $z=t$ and $z=0$ can be worked out from their asymptotic behavior around $z\sim 0$, $\psi_\theta(z)\sim z^{\frac{1+\theta\sqrt{1-4\delta}}{2}}$, as 
\begin{equation}
    M=\lt(\begin{array}{cc}
       e^{i\pi\lt(1+\sqrt{1-4\delta}\rt)}  &  \\
         & e^{i\pi\lt(1-\sqrt{1-4\delta}\rt)}\\
    \end{array}\rt).
\end{equation}
When $\Delta\gg c$, i.e. $\delta\gg 1$, one can solve for $\psi_\pm$ with the WKB approximation by noting that $C_2\sim\Delta\gg c$ and the last term in \eqref{eq:BPZ} dominates the potential, 
\begin{equation}
    \psi_\pm\sim \exp\lt[\pm i\sqrt{t(1-t)C_2}\int^z\frac{{\rm d}z'}{\sqrt{z'(z'-1)(z'-t)}}\rt].
\end{equation}
Comparing it with the monodromy matrix, we have at the leading order of $\Delta/c$, 
\begin{align}
    \sqrt{t(1-t)C_2}\int_0^t\frac{{\rm d}z'}{\sqrt{z'(z'-1)(z'-t)}}=2\pi i\sqrt{\delta}+{\cal O}(1).
\end{align}
Note here that the integrand has a branch cut between $z=0$ and $z=t$, and therefore the contour integral circling both $z=0$ and $z=t$ reduces to twice the integral slightly above the branch cut. 
Using the following integral representation of the hypergeometric function, 
\begin{align}
    &z^{1-c}\ _2F_1[1+b-c,1+a-c,2-c,z]\cr
    &=\frac{\Gamma(2-c)}{\Gamma(1+b-c)\Gamma(1-b)}\int_0^z{\rm d}s\ s^{b-c}(1-s)^{c-a-1}(z-s)^{-b},
\end{align}
we obtain 
\begin{equation}
    C_2=-\frac{\pi^2\delta}{t(1-t)K^2(t)},
\end{equation}
with 
\begin{equation}
    K(t):=\frac{1}{2}\int_0^t{\rm d}s\ s^{-\frac{1}{2}}(1-s)^{-\frac{1}{2}}(t-s)^{-\frac{1}{2}}=\frac{\pi}{2}\ _2F_1\lt[\frac{1}{2},\frac{1}{2},1,t\rt].
\end{equation}
One can integrate $C_2$ over $t$ to find ${\cal F}_{cl}$ as 
\begin{equation}
    {\cal F}_{cl}=\pi \delta\frac{K(1-t)}{K(t)}+{\cal O}(1).
\end{equation}

In our discussion, $C_2\to u$ in the semiclassical limit and we wish to find $\delta\to a^2$ at the leading order ${\cal O}(r_h^{-2})$. From the map \eqref{map-brane}, we see that $\delta_1$, $\delta_t$, $\delta_\infty$ and $C_2$ are dominant in the potential, and thus the WKB approximation gives 
\begin{equation}
    \psi_\pm \sim \exp\lt[\pm\frac{\omega}{2r_h^2}\int^z{\rm d}z'\frac{1}{(1-z')(z'+1)}\sqrt{\frac{2(2z^{\prime 3}+(1-v)z^{\prime 2}+v-1)}{z'}}\rt],
\end{equation}
where we set $t=-1$ and substituted the semiclassical result $C_2\to u=:\frac{\omega^2}{4r_h^4}v$. $v=1+{\cal O}(r_h^4)$ at the leading order. Unlike the previous case, there is no branch cut between $z=0$ and $z=t$, but instead, there is a pole at $z=t=-1$. The monodromy can then be computed from the residue at $z=-1$, 
\begin{equation}
    \frac{\pi i\omega}{r_h^2}=2\pi i\sqrt{\delta}+{\cal O}(1).
\end{equation}
This suggests that 
\begin{equation}
    a^2=-\frac{\omega^2}{4r_h^4}+{\cal O}(1),
\end{equation}
and proves the necessary condition \eqref{nec-cond}.

\bibliographystyle{JHEP}
\bibliography{Lifshitz}

\end{document}